\documentclass{aastex62}

\usepackage{float}
\usepackage{bigstrut}
\usepackage{hhline}
\usepackage{lipsum}

\newcommand{\ikt}{{\it Kepler}}
\newcommand{\ik}{{\it Kepler~}}
\renewcommand{\arraystretch}{1.5}
\received{26 August 2020}
\revised{4 April 2021}
\accepted{7 April 2021}
\submitjournal{ Icarus}

\shorttitle{Orbital stability of compact three-planet systems}
\shortauthors{Lissauer \& Gavino}

\begin{document}

\title{Orbital stability of compact three-planet systems, I: Dependence of system lifetimes on initial orbital separations and longitudes }

\correspondingauthor{Jack J. Lissauer}
\email{ Jack.Lissauer@nasa.gov }

\author{Jack J. Lissauer}
\affiliation{Space Science \& Astrobiology Division\\
MS 245-3\\
NASA Ames Research Center \\
 Moffett Field, CA 94035, USA}

\author{Sacha Gavino}\footnote{Sacha Gavino's current address is Niels Bohr Institute and Centre for Star and Planet Formation, University of Copenhagen, \O{}ster Voldgade 5-7, 1350, Copenhagen K, Denmark}
\affil{Laboratoire d'Astrophysique de Bordeaux, Universit\'e de Bordeaux, CNRS, F-33615\\
Pessac, France}



\begin{abstract}

We explore the orbital dynamics of systems consisting of three
planets, each as massive as the Earth, on coplanar, initially circular,  orbits about a star of one solar mass. The initial semimajor axes of the planets are equally spaced in terms
of their mutual Hill radius, which is equivalent to a geometric progression of orbital periods for small planets of equal mass. Our simulations
explore a wide range of spacings of the planets, and were integrated
for virtual
times of up to 10 billion years or until the orbits of any pair of planets crossed.   We find the same general trend of system lifetimes increasing exponentially with separation between orbits seen by previous studies of systems of three or more planets. One focus of this paper is to go beyond the rough trends found by previous numerical studies and quantitatively explore the nature of the scatter in lifetimes and  the destabilizing effects of mean motion resonances. In contrast to previous results for five-planet systems, a nontrivial fraction of three-planet systems survive at least several orders of magnitude longer than most other systems with similar initial separation between orbits,  with some surviving $10^{10}$ years at much smaller orbital separations than any found for five-planet systems. Substantial shifts in the initial planetary longitudes cause a scatter of roughly a factor of two in system lifetime, whereas the shift of one planet's initial position by 100 meters along its orbit results in smaller changes in the logarithm of the time to orbit crossing, especially for systems with short lifetimes.

\end{abstract}



\section{Introduction} \label{sec:intro}

NASA's \ik mission has discovered hundreds of multiple planet systems \citep{Lissauer:2014, Rowe:2014}. Many of these multi-planet systems are quite tightly packed, with adjacent planets' orbits much closer than the orbits of our Solar System's planets, both in physical distance and dynamically \citep{Lissauer:2011a, Fabrycky:2014}. For example, the Kepler-11 system
harbors six known planets, five of which have orbital periods between 10 and 47 days \citep{Lissauer:2011b}. Studying the orbital stability of such tightly-packed multiple planet systems provides clues to how they
form and how long they survive. 

The stability of systems of two planets has been characterized analytically  \citep{Hill:1878a, Hill:1878b, Hill:1878c, Gladman:1993}. In contrast,  analytic studies have not provided as comprehensive a solution to delineating stability boundaries for systems with more than two planets, so such systems have been studied extensively using numerical integrations (e.g.,  \citet{Chambers:1996}, \citet{Marzari:2002}, \citet{Smith:2009}, \citet{Pu:2015},
\citet{Morrison:2016}, \citet{Tamayo:2016}, \citet{Obertas:2017}). \citet{Lissauer:1995} examined the spacing of planets and moons within the Solar System and proposed approximate criteria for global stability of more than two planets based upon resonance overlapping and the Hill/Jacobi exclusion zone.  

The present study can be seen as an extension of \citet{Smith:2009} (henceforth SL09), who investigated
the stability of closely-spaced three-planet systems with the same masses and relative spacings as we use herein. The main goal of this
paper is to perform a much more extensive study of the stability timescale of three-planet systems  that are uniformly spaced in units of  mutual Hill radii to better survey their diversity in lifetime, considering the influence of the initial planetary longitudes as well as (the previously-varied parameter) orbital separation. We compare the lifetimes of the systems that we study with the lifetimes of analogous five planet systems presented by \citet{Obertas:2017} (henceforth OVT17) and those of three-planet systems orbiting $\alpha$ Centauri A or B  \citep{Quarles:2018}.   For this purpose,  we have integrated $\sim$~18,000 systems, each consisting of three Earth-like planets orbiting a solar mass star, until a pair of planetary orbits crossed or a pre-selected interval of $10^8$ or $10^{10}$ virtual years elapsed.  Throughout this work, we quote values of the initial separations between the planets' orbits in terms of mutual Hill sphere radii (Eq.~\ref{eq:Hill}). Nonetheless, since the masses of the planets and the star are identical for all of the simulations presented herein, the listed planetary separations can easily be scaled to values relative to the critical separation for two-planet systems defined by the resonance overlap criterion \citep{Chirikov:1979, Wisdom:1980, Deck:2013, Hadden:2018} and for overlapping of three-body resonances in systems of three planets \citep{Quillen:2011, Petit:2020}.  

In Section \ref{sec:methods}, we present the  methodology used to perform our integrations. An analysis of the lifetimes of a set of simulations with initial longitudes of the outer two planets each chosen randomly and independently, which we performed to provide a direct comparison of the lifetimes of three-planet systems with those of analogous five-planet systems simulated by OVT17, is given in \S\ref{sec:OVT}.  We present exponential fits to the lifetimes of these systems as a function of initial orbital separation in \S \ref{sec:exponentialOVT}; deviations from the exponential trend are analyzed in \S\S \ref{sec:deviationsOVT} and \ref{sec:deviations}. Section \ref{sec:results} shows the results of the primary set of simulations that we performed, which were begun with the planets approaching conjunction. In \S \ref{sec:resonances}, we explain the displacement of short-lifetime regions inwards of first-order resonances found by OVT17.  Lifetimes of planetary systems with the same combination of initial angles used by SL09 are presented and analyzed in \S \ref{sec:SL09}. Section  \ref{sec:chaos} analyzes the effects of  very small changes in initial longitudes on lifetimes. We present additional runs testing the effects of large changes in initial longitudes of the planets  on system lifetimes in \S \ref{sec:altlong}.  We summarize and discuss our results in Section \ref{sec:discussion} and compare our findings with those of previous numerical studies in Section \ref{sec:compare}.  Section \ref{sec:epi} is an epilogue discussing departures from the Newtonian dynamics of point-mass bodies that must be considered when applying our results to planets on very short-period orbits.

\section{Methods} \label{sec:methods}

We use numerical integrations to study the stability of systems consisting of three planets, each with mass equal to that of the Earth, 1 M$_\oplus$, orbiting a one solar mass, 1 M$_\odot$, star.  In all cases, the systems are coplanar, the planets orbit the star in the same direction, the orbits of the planets are initially circular, and the inner planet initially orbits at 1 AU from the star. The problem is modeled using purely Newtonian dynamics, so provided time is measured in units of the initial orbital period of the inner planet, the equations and therefore the results would be unchanged if the masses of the star and planets were scaled up or down proportionately, keeping the mass ratios constant, and/or if the planets were all moved closer to or farther from the star, as long as the distance ratios were kept constant and speeds were adjusted to keep the initial orbits circular.   Our integrations are performed using heliocentric coordinates; because the planets/star mass ratio is quite small, the difference from Jacobi coordinates is slight.

\subsection{Initial Semimajor Axes of the Planets} \label{sec:a0}

We measure the initial orbital separations using the unit of mutual Hill radius of neighboring planets, $R_{H_{j, j+1}}$, where $j = 1$ refers to the inner planet,  $j = 2$  to the middle planet, and  $j = 3$  to the outer planet. The mutual Hill radius is given by:

\begin{equation}
\label{eq:Hill}
	R_{H_{j, j+1}} = \left[\frac{m_j + m_{j+1}}{3\tilde{M_j}}\right]^\frac{1}{3}\frac{\left(a_j + a_{j+1}\right)}{2} = \left[\frac{2{\rm M}_\oplus}{{3({\rm M}_\odot}+(j-1){\rm M}_\oplus)} \right]^\frac{1}{3} \frac{\left(a_j + a_{j+1}\right)}{2} ,
\end{equation} 
 
\noindent where $a_j$ and $m_j$ are the semimajor axis and the mass of the $j^{th}$ planet, respectively, and we define

\begin{equation}
\label{eq:SumMass}
	\tilde{M}_j \equiv M_\star + \sum_{k=1}^{j-1} m_k = {\rm M}_\odot + (j-1){\rm M}_\oplus,
\end{equation} 

\noindent where $M_\star$ is the mass of the host star. 
The first equalities in Eqs.~(\ref{eq:Hill}) and (\ref{eq:SumMass}) give the general formulas for any stellar and planetary masses, whereas the second parts are applicable specifically to the systems studied herein.  
\noindent We use a dimensionless number, $\beta$,  to specify the initial spacing of adjacent planetary orbits:

\begin{equation}
\label{eq:beta}
	\beta \equiv \frac{a_{j+1} - a_j}{R_{H_{j, j+1}}} .
\end{equation} 

\noindent Equation (\ref{eq:Hill}) can be rearranged to give:

\begin{equation}
\label{eq:developed}
	a_{j+1} = a_j + {\beta}R_{H_{j, j+1}} = a_j + {\beta}{\left[\frac{m_j + m_{j+1}}{3\tilde{M}_j}\right]}^{\frac{1}{3}}\frac{(a_j + a_{j+1})}{2} .
\end{equation}

\noindent Solving Eq.~(\ref{eq:developed}) for $a_{j+1}$ yields:

\begin{eqnarray}
\label{eq:solved}
	a_{j+1} = {}a_j\left[1 + \frac{\beta}{2}{\left(\frac{m_j + m_{j+1}}{3\tilde{M}_j}\right)}^{\frac{1}{3}}\right]{\left[1 - \frac{\beta}{2}{\left(\frac{m_j + m_{j+1}}{3\tilde{M}_j}\right)}^{\frac{1}{3}}\right]}^{-1} \nonumber\\ 
	 = a_j\left[1 + \beta\left(\frac{{\rm M}_\oplus}{12({\rm M}_\odot+(j-1){\rm M}_\oplus)} \right)^\frac{1}{3}\right]\left[1 - \beta\left(\frac{{\rm M}_\oplus}{12({\rm M}_\odot+(j-1){\rm M}_\oplus)} \right)^\frac{1}{3}\right]^{-1}.
\end{eqnarray}

\noindent The first equality in Eq.~(\ref{eq:solved}) gives the general formula for any stellar and planetary masses, whereas the second part is applicable specifically to the systems studied herein.  We used Eq.~(\ref{eq:solved}) to determine, based upon the spacing parameter $\beta$ and the Hill radius, the initial semimajor axes of all planets in our simulated systems\footnote{Our Eq.~(\ref{eq:solved}) is equivalent to Eq.~(\ref{eq:developed}) of SL09. We discovered that the expanded version of this formula given in Eq.~(\ref{eq:solved}) of SL09 contains a misprint. The correct form is given in Eq.~(\ref{eq:solved}) of \citet{Smith:2010}. The calculations in both Smith and Lissauer papers used the correct formula.}. 

An initially circular and coplanar two-planet system will remain stable forever if the initial Hill separation  $\beta \geq 2\sqrt{3} \approx 3.4641$. (If, in analogy to the simulations presented in this article, both of the planets have masses equal to that of Earth, the star's mass is 1 M$_\odot$ and the inner planet orbits at 1 AU, then the critical semimajor axis for the outer planet is $a_2 \approx 1.04464$ AU, which implies an initial orbital period of 1.0677 years.)   For this reason we do not compute any systems for which $\beta < 2\sqrt{3}$.  


\subsection{Initial Longitudes of the Planets} \label{sec:theta0}

Following OVT17, we drew the initial longitudes for the middle and outer planets randomly and independently\footnote{OVT17 also drew the initial longitude of the inner planet in their study, but since only the relative longitudes of the planets matter, we placed the inital longitude of the inner planet at 0 for simplicity.} for each of the simulations presented in Section \ref{sec:OVT}.  To separate the effects of variations in initial longitudes from those caused by differences in the initial distances between the orbits, the same initial planetary longitudes are used for all of the simulations presented within each of  Sections \ref{sec:results}, \ref{sec:SL09}, \ref{sec:altlong} and  \ref{sec:aligned}, and a very small range in initial longitudes is examined in Section \ref{sec:chaos}.  

For our Primary set of runs, results of which are presented in Section \ref{sec:results}, the initial mean anomaly\footnote{As the orbits initially circular, this is also the starting value of the true anomaly, i.e., the initial longitude, $\theta$.} of the $j^{th}$ planet is $2{\pi}(j-1)\lambda$ degrees, where 

\begin{equation}
\label{eq:Euler}
	\lambda \equiv \frac{1+\sqrt{5}}{2} = 1.618...
\end{equation}

\noindent is Euler's number (a.k.a.~the golden ratio). Simulations presented in Section \ref{sec:SL09} use the same initial longitudes employed by SL09, where the initial mean anomaly of the $j^{th}$ planet is $2{\pi}(j-1)\lambda$ radians; SL09 chose this initial longitudinal separation in order to avoid choosing special orientations (i.e., conjunction, opposition, etc.) regardless of the number of planets simulated. Section \ref{sec:chaos} investigates the effects of chaos by moving the starting position of one planet along its orbit by 100 meters relative to the longitudes in our SL09 simulations.
In Section \ref{sec:altlong}, we present crossing times from simulations of system with initial planetary longitudes that are three of the vertexes of a regular hexagon and compare them with those presented in Section \ref{sec:results}.  Results for systems started with all planets at the same longitude (Aligned) are given in Section \ref{sec:aligned}. 

\begin{table}[htbp]
\caption{ Initial Planetary Longitudes}
\begin{center}
\begin{tabular}{| l  r || c | c | c | }
\hline
\textbf{Set(s) of runs} &&  \multicolumn{1}{c|}{\textbf{Section(s)}} & \multicolumn{1}{c|}{\textbf{$\theta_{2} - \theta_{1}$}} & \multicolumn{1}{c|}{\textbf{$\theta_{3} - \theta_{1}$}}  \bigstrut \\ \hline
Random && \ref{sec:OVT} & random & random  \\ \hline
Primary && \ref{sec:results}, \ref{sec:resonances} & 10.17$^\circ$ & 20.33$^\circ$   \\ \hline
SL09, SL09$^\dagger$, Chaos  &&  \ref{sec:SL09},  \ref{sec:chaos} & 222.49$^\circ$ & 84.98$^\circ$  \\ \hline
Hexagonal &&  \ref{sec:altlong} & 180$^\circ$ & 60$^\circ$ \\ \hline
Aligned &&  \ref{sec:aligned} & 0$^\circ$ & 0$^\circ$ \\ \hline
\end{tabular}
 \end{center}
   \tablecomments{ \label{tab:initial} Initial longitudes of middle and outer planets ($\theta_{2}$ and  $\theta_{3}$, respectively) relative to that of the inner planet ($\theta_{1}$), measured along the direction of the orbits, for each set of runs presented herein. Values presented in this table have been rounded to the nearest $0.01^\circ$. The initial longitudes of the SL09, SL09$^\dagger$ and Chaos sets differ by $\ll 0.01^\circ$. }
 \end{table}
 
\begin{figure}[H]
\centering
\includegraphics[scale=0.5]{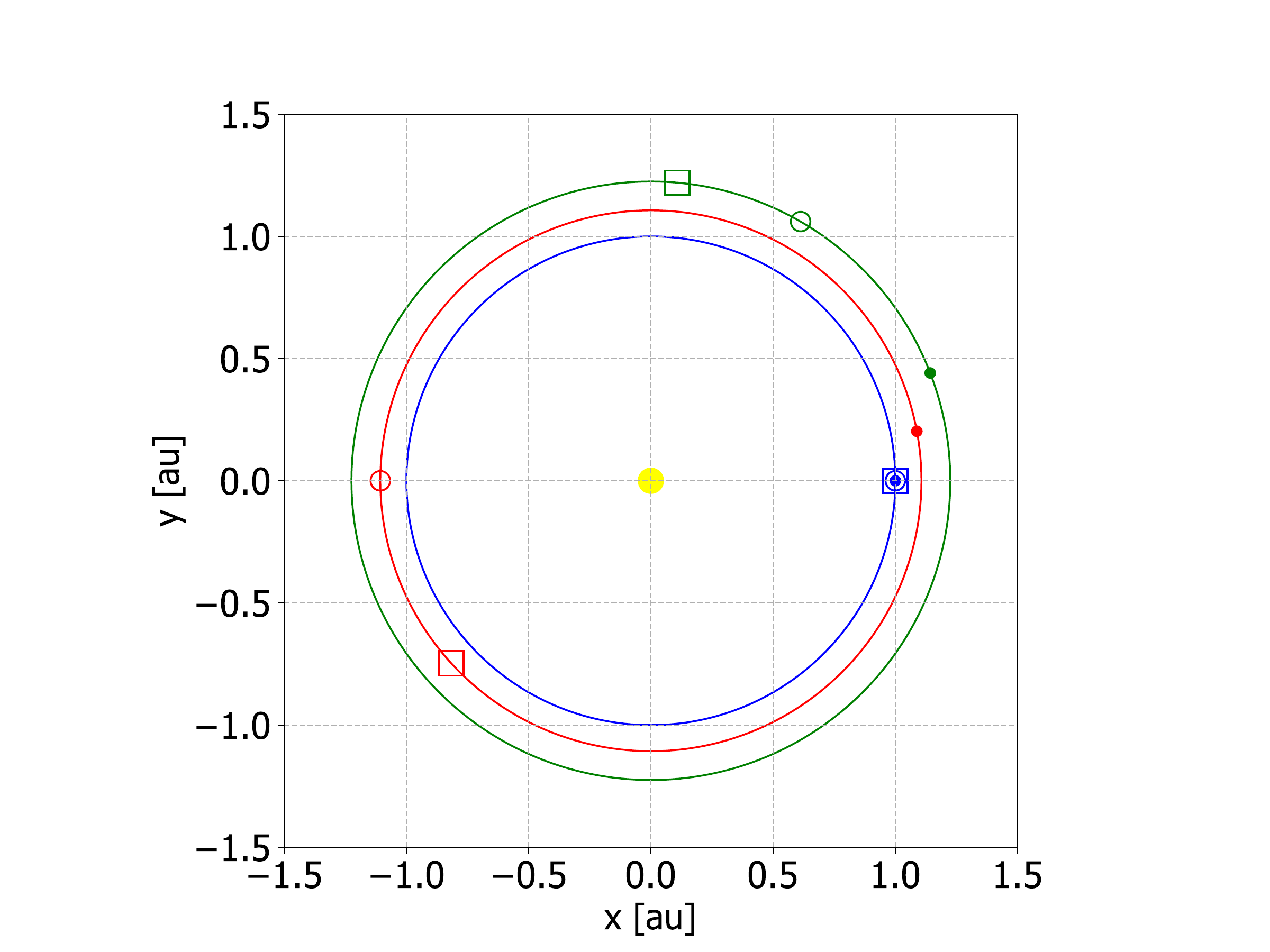}
\caption{Initial orbits and longitudes for dynamical separation $\beta = 8$. All planets orbit in the counterclockwise direction. The filled circles show the planets at their initial position for the Primary set of runs, the open squares represent the initial positions of the planets for the SL09 runs, while the open circles show planets at their initial position for the Hexagonal longitude set of runs presented in Section \ref{sec:altlong}. } 
\end{figure}

\subsection{Integrator and Duration of the Simulations}\label{sec:integrator} 
  
 The Wisdom-Holman method MVS (mixed-variable symplectic) integrator in the {\tt Mercury} software package (Chambers 1998), with a timestep of 18 days, was used to perform all of the integrations presented in this manuscript. Integrations presented in Section \ref{sec:chaos} that study the effects of very small alterations in initial orbital longitudes on system lifetimes use a minor modification to {\tt Mercury} provided to us by John Chambers (private communication, 2020) that more accurately handles the input of initial orbital longitudes of the planets\footnote{The problem occurred because  sin$^2x$ + cos$^2x$ doesn't always equal 1, at least not when $x$ is small and precision is finite. To fix it, we edited the subroutine mco$\_$sine in mercury6$\_$5.for and in element6.for.  The existing lines, starting with pi = ..., and ending with the last end if statement, were commented out. Then the following two lines were added: (1) ``cx = cos(x)'', and (2)  ``sx = sin(x)''.}; this modification does not, to the extent tested, systematically affect lifetimes of systems (Table \ref{tab:correction}). 
 
 We integrate each system until two orbits cross (the apoapsis of the innermost orbit exceeds the periapsis of the middle orbit or the apoapsis of the middle orbit exceeds the periapsis of the outermost orbit) or until a specified amount of time, equal to either $10^8$ or $10^{10}$ times the initial orbital period of the innermost planet, has elapsed.  We refer to this orbital crossing time as the lifetime of the system or the time for the system to become unstable, and denote it by $t_c$.  We quote values of $t_c$ normalized by the initial period of the innermost planet (i.e., in years).

Our code tests for orbit crossing once every year for nondimensional orbital separation $\beta < 5$ and once every $\sim$~10 years for $\beta \ge 5$.  In some systems, orbits may have crossed briefly but returned to nested positions prior to the next check, so very short-lived episodes of orbit crossing may be missed.  Excluding very short-lived systems, most of the time in the integration is typically spent waiting
for small variations in eccentricity to diffusively build up and allow stronger interactions between the planets, so we expect that this rarely leads to substantial overestimates of system lifetimes, especially if there is an actual close encounter between two planets.  However, our procedure does quantize estimated values of $t_c$ by rounding  actual crossing times upwards by up to $\sim$~10 years.

\section{Results: Random Initial Longitudes}\label{sec:OVT}

For the integrations presented in this section, we select values of the initial longitudes of the middle and outer planets, $\theta_2$ and $\theta_3$,  for each system randomly and independently from a uniform distribution in angle, as was done by \citet{Obertas:2017}. As in OVT17 and throughout this paper, the same value of $\beta$ (denoted as $\Delta$ by OVT17) was used for the inner and outer pairs of planets.

We simulated systems for all separations $\beta$ that are multiples of 0.0005 within the range [3.4645, 6.73], and multiples of 0.01  within the ranges (6.73, 9], [9.5, 9.61] and [10.53, 10.67].  We stopped each integration if two orbits crossed or when the simulation time reached a limit of $10^8$ years ($10^{10}$ years for $\beta < 5.98$). 

\begin{figure}[H]
\centering
\includegraphics[scale=0.75]{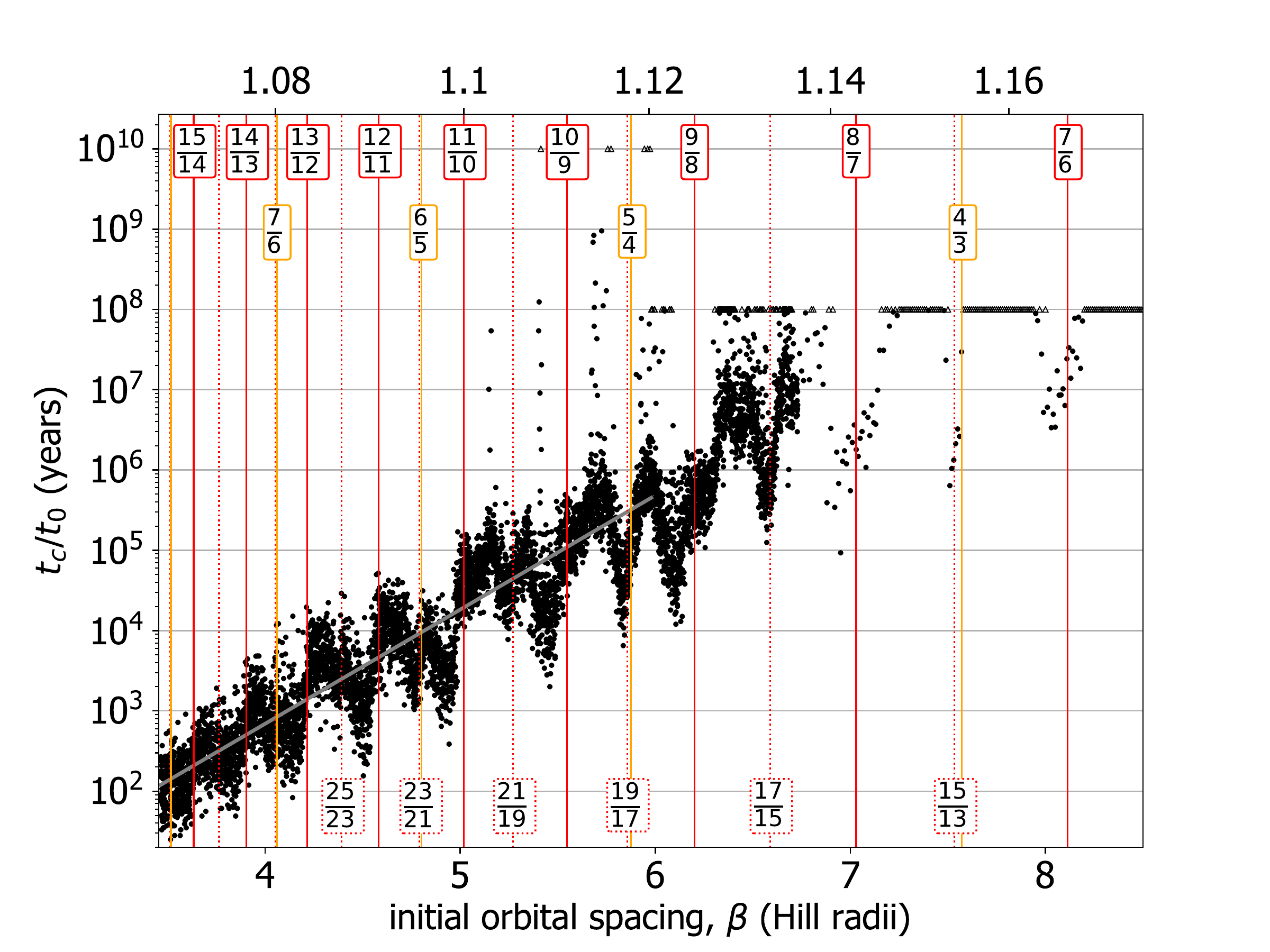}
\caption{Lifetime, $t_c$, of three-planet systems as a function of the initial separation of the orbits of neighboring planets in units of mutual Hill sphere radii, $\beta$ (Eq.~\ref{eq:beta}), for our Set of runs with initial longitudes of the middle and outer planets selected randomly and independently for each system. The open triangles represent systems that remained stable for the entire time interval, either $10^8$ or   $10^{10}$ years, simulated. The systems that survived for ten billion years had initial orbital separations of $\beta = 5.4135$, 5.7565, 5.7730, 5.9445, 5.961 and 5.979.  The top axis gives the initial period ratio of neighboring planets; note that the scale of the plot is not (quite) linear in this variable. The solid red vertical lines denote the locations of first-order mean motion resonances (see \S\ref{sec:resonances} for definitions) between adjacent pairs of planets, the  dotted red vertical lines denote the locations of second-order MMRs between adjacent pairs of planets, and the solid gold vertical lines denote first-order resonances between the innermost and outermost planets. The solid gray line segment represents the fit to all points within the range [3.4645,  5.9795]. Note the narrow upward ``spikes'' in the lifetimes of some systems near $\beta = 5.15$ and 5.41, and the wider spikes for separations near $\beta = 5.7$ and 6.
} 
\label{fig:random}
\end{figure}

Figure \ref{fig:random} shows the crossing time, $t_c$, as a function of initial orbital separations given in mutual Hill radii up to $\beta = 8.5$. The overall trend is that more widely spaced systems tend to be longer-lived, but the data show significant scatter about the general trend.  We expect that the choice of relative initial longitudes of the planets, mean motion resonances (MMRs) and   chaos  all contribute to this scatter; we analyze these three affects in \S\ref{sec:results}, \S\ref{sec:resonances} and \S\ref{sec:chaos}, respectively. 

Of the more widely-spaced systems studied, only two in the range [8.51, 9] went unstable during the $10^8$ years simulated -- the $\beta$ values of these systems were 8.78 and 8.79, placing them slightly inward of the 13:11 MMR of neighboring planets, and even these systems survived for more than $6 \times 10^7$ years. The intervals simulated for separations $\beta > 9$ were selected to cover the regions near and slightly inwards of the 6:5 MMR of neighboring planets, as well as similar regions in the vicinity of the 11:9 MMR of neighboring planets and the 3:2 MMR of the inner and outer planets, respectively. None of the   runs with separations $\beta > 9$  became unstable during the 10$^8$  orbits integrated.

Comparing our Fig.~\ref{fig:random} with Figures 2 and 3 of OVT17 confirms similarities in trends of lifetime vs.~orbital separation for systems of three and five planets, apart from the tendency of three-planet systems to be longer-lived at a given separation (\citealt{Chambers:1996}, SL09).   
Although the spread in lifetimes of most three-planet systems is  similar to that present in the OVT17 data for five-planet systems, a small but not insignificant fraction of three-planet systems survive for orders of magnitude longer than most systems having similar initial orbital separations.  These exceptionally long-lived systems are only found over limited regions in orbital separation, $\beta$, and are discussed further in Sections \ref{sec:deviationsOVT}, \ref{sec:deviations} and \ref{sec:SL09} -- \ref{sec:discussion}.
     
\subsection{Lifetime vs.~Separation: Exponential fits}\label{sec:exponentialOVT}

Previous studies have found that $\log t_c$ of systems of three or more planets increases approximately linearly with the initial separation measured  in mutual Hill radii \citep{Chambers:1996, Duncan:1997, Faber:2007, Smith:2009, Morrison:2016, Obertas:2017}, according to the following relationship:

\begin{equation}
\label{eq:exponential}
	\log{t_c} = b{\beta} + c ,
\end{equation}
 
\noindent where $b$ and $c$ are constants. We used the durations of the simulations in place of crossing times for systems that survived; this introduces a downward bias in the fit at high values of $\beta$ and thus reductions in the value of $b$, but by smaller amounts than if the surviving systems were omitted. To reduce the amount of bias, fits were not performed over regions in which integrations of stable systems were halted prior to $10^{10}$ years.

We fit our data over the interval $3.4645 \leq \beta \leq$ 5.9795 to  the form of Equation (\ref{eq:exponential}) and computed values of the constants to be $b = 1.4220$ and $c = -2.8482$. Extrapolation of this exponential fit to more widely-spaced systems estimates the separation at which $t_c = 10^{10}$ years to be $\beta=9.0921$.  The analogous coefficients for five-planet systems found by OVT17 were $b = 0.951$ and $c = -1.202$ (their fit covered the range $2\sqrt3 \le \beta \le 8.4$). These coefficients imply similar lifetimes for systems of three and five planets with $\beta \approx 2\sqrt3$, but that the lifetimes of three-planet systems characteristically increase faster with orbital separation than do those of five-planet systems. 

Because the fit data all have $\beta$ values well above 0, the derived value of the constant $c$, which corresponds to the value of the fit line at $\beta = 0$, depends strongly on the value of the slope, $b$. To reduce this correlation, following \citet{Quarles:2018}, we  performed fits to the shifted exponential

\begin{equation}
\label{eq:shiftexponential}
	\log{t_c} = b'{\beta'} + c', 
\end{equation}

\noindent where $\beta' \equiv \beta - 2\sqrt{3}$. 

Table \ref{tab:fit} shows results\footnote{ Note that these values, as well as all other results presented herein, apply only to system of three planets with equal masses that orbit a star whose mass ratio to each of the planets is $M_\odot/M_\oplus$.} from fitting Eq.~(\ref{eq:shiftexponential}) to various subsets of our crossing times, including data from Sets of runs with identical starting longitudes (Sections \ref{sec:results} and \ref{sec:SL09} --  \ref{sec:aligned}) as well as from the independent random draw longitudes Set displayed in Fig.~\ref{fig:random}. In addition to quoting values of the coefficients $b'$ and $c'$  and the standard deviation of points from the fit line, $\sigma_\mathrm{exp}$, we give the value of $\beta$ (not $\beta'$, which is less familiar) at which the exponential fit predicts $t_c = 10^{10}$ years  and three other measures of dispersion of the crossing times that are described in \S \ref{sec:distdev}. This shifted fit returned the coefficients listed in the third numerical row of Table \ref{tab:fit} for the same data fit above; coefficients for other ranges in $\beta$ are listed in the first and second rows.  Note that the vast majority of the runs used in the fits presented in  Table \ref{tab:fit}  ended in orbit crossing, and all of the exceptions were run for $10^{10}$ years. The number of stopped runs in the data set used for each fit   is also provided in this table.

  \begin{table}[htbp]
\caption{ Coefficients of Log-Linear Fits to System Lifetimes}
\begin{center}
\footnotesize
\begin{tabular}{| l | l | l || c | c | c | c | c | c | c | c |}
\hline
Set of runs & Range &  \multicolumn{1}{l ||}{Resol'n} & \multicolumn{1}{c|}{ Slope, $b'$} & \multicolumn{1}{c|}{Intercept, $c'$} & \multicolumn{1}{c|}{$\beta$ at $\log t_c = 10$} & \multicolumn{1}{c|}{$t_c >10^{10}$} &  \multicolumn{1}{c|}{\textbf{$\sigma_\mathrm{exp}$}} &  \multicolumn{1}{c|}{\textbf{$\sigma_\mathrm{local}$}} &   \multicolumn{1}{c|}{\textbf{$\sigma_\mathrm{exp}^\mathrm{cum}$}} & \multicolumn{1}{c|}{\textbf{$\sigma_\mathrm{local}^\mathrm{cum}$}}   \bigstrut \\ \hhline{|=|=|=||=|=|=|=|=|=|=|=|}
Random & 3.4645 -- 5.1870 & 0.0005 & 1.474 $\pm$ 0.014 & 2.046 $\pm$ 0.014 & 8.862 $\pm$  0.044 & 0 &   0.420 &   0.288 & 0.414 & 0.261 \\ \hline 
Random & 3.4645 -- 5.9100 & 0.0005 & 1.396 $\pm$ 0.010 & 2.099 $\pm$ 0.014 & 9.122 $\pm$ 0.031 & 3 & 0.484 & 0.352 &  0.436  & 0.267\\ \hline
Random & 3.4645 -- 5.9795 & 0.0005 & 1.422 $\pm$ 0.010 & 2.078 $\pm$ 0.014 & 9.035 $\pm$ 0.029  & 6 &0.497& 0.368& 0.434 &0.268 \\ \hhline{|=|=|=||=|=|=|=|=|=|=|=|}

Primary & 3.465 -- 5.187 & 0.001 & 1.468 $\pm$ 0.019  & 2.108 $\pm$ 0.019& 8.840 $\pm$ 0.059 & 0  &0.420 & 0.280 &  0.393 &  0.225\\ \hline
Primary & 3.465 -- 5.910 & 0.001 & 1.399 $\pm$ 0.014  & 2.163 $\pm$ 0.020  & 9.066 $\pm$ 0.044 & 1 & 0.507 & 0.349 & 0.416 & 0.234 \\ \hline
Primary & 3.47 -- 6.30 & 0.01 & 1.603 $\pm$ 0.070  & 2.026 $\pm$ 0.114  & 8.439 $\pm$ 0.152 & 8 & 0.955 &0.664&  0.549 &   0.319\\ \hline
Primary & 3.47 -- 6.75  &  0.01 & 1.556 $\pm$ 0.054 & 2.071 $\pm$ 0.103 & 8.560 $\pm$ 0.117 & 9  & 0.925 & 0.629 & 0.526 & 0.324 \\ \hline
Primary & 3.5 -- 7.5  &  0.1 & 1.680 $\pm$ 0.128 & 1.799 $\pm$ 0.300 & 8.346 $\pm$ 0.207 & 5 & -- & -- & -- & -- \\ \hhline{|=|=|=||=|=|=|=|=|=|=|=|}

 SL09& 3.465 -- 5.187 & 0.001 &  1.495 $\pm$ 0.020 & 1.968 $\pm$ 0.020  & 8.838 $\pm$ 0.060  &  0 &  0.414 & 0.246 &  0.394 &  0.236 \\ \hline   
SL09 & 3.465 -- 5.910 & 0.001 & 1.492 $\pm$ 0.015 & 1.967 $\pm$ 0.021 & 8.847 $\pm$ 0.042 & 0 & 0.528 &  0.297 &  0.434 &  0.254 \\ \hline   
SL09 & 3.465 -- 6.300  &  0.001 & 1.383 $\pm$ 0.012 & 2.064 $\pm$ 0.020 & 9.201 $\pm$ 0.038 & 0 &0.540 &  0.300 &  0.447 &  0.258 \\ \hline  
SL09 & 3.47 -- 7.15  &  0.01 & 1.478 $\pm$ 0.038 &  1.977 $\pm$ 0.081 & 8.892 $\pm$  0.090 &5 & 0.733 &  0.375 &  0.574 & 0.306 \\ \hline  
SL09 & 3.5 -- 7.5  &  0.05 & 1.621 $\pm$ 0.076 & 1.808 $\pm$ 0.179 &  8.512 $\pm$ 0.136 & 3 &-- & -- & -- & -- \\ \hhline{|=|=|=||=|=|=|=|=|=|=|=|}

SL09$^\dagger$ & 3.47 -- 6.32  &  0.01 & 1.398 $\pm$ 0.037  & 2.027 $\pm$ 0.062 & 9.166 $\pm$ 0.113 & 0 &0.519 &  0.286 &  0.460 &  0.257 \\ \hline  
Chaos & 3.47 -- 6.32  &  0.01 & 1.406 $\pm$ 0.037  & 2.028 $\pm$ 0.060 & 9.132 $\pm$ 0.108 & 0 & 0.507 & 0.325 &  0.438 & 0.273 \\ \hhline{|=|=|=||=|=|=|=|=|=|=|=|}
 Hexagonal &  3.465 -- 5.187 & 0.001 & 1.405 $\pm$ 0.019 &  2.106 $\pm$ 0.019 &  9.084 $\pm$ 0.065 &  0 &  0.401 & 0.240 & 0.395 &  0.224 \\ \hline
Hexagonal & 3.465 -- 5.910  &   0.001 & 1.344 $\pm$ 0.012 & 2.150 $\pm$ 0.017 & 9.310 $\pm$ 0.133 & 0 &0.418 &  0.254 &  0.412 &  0.235 \\ \hline
Hexagonal & 3.47 -- 5.94  &  0.01 & 1.337 $\pm$ 0.038 & 2.166 $\pm$ 0.054 & 9.323 $\pm$ 0.129 & 0 &0.421 & 0.277 & 0.411 &  0.253 \\ \hhline{|=|=|=||=|=|=|=|=|=|=|=|}
Aligned & 3.465 -- 5.187 & 0.001 & 1.461 $\pm$ 0.024 & 2.348 $\pm$ 0.024 & 8.702 $\pm$ 0.072 & 2 & 0.497 &  0.325 & 0.410 &  0.229 \\ \hline
\end{tabular}
 \end{center}          
   \tablecomments{Values of the  coefficients computed from the exponential fits to  the lifetimes of more than ten thousand systems in the various Sets of integrations studied herein. From left to right, the columns in this table list the name of the Set, range  of initial orbital separation, resolution in $\beta$, the derived coefficients $b'$ and $c'$, the value of $\beta$ where the orbital crossing time predicted by the fit line is $10^{10}$ years,  the number of runs that survived for the entire $10^{10}$ years simulated, and two measures of the dispersions of the data relative to both the exponential fit, $\sigma_\mathrm{exp}$ and $\sigma_\mathrm{exp}^\mathrm{cum}$, and to the running median, $\sigma_\mathrm{local}$ and $\sigma_\mathrm{local}^\mathrm{cum}$ (see \S\ref{sec:distdev} for details).}  \label{tab:fit}  
 \end{table}

\subsection{Deviations from the Exponential Fits}\label{sec:deviationsOVT}

Figure \ref{fig:random} shows that system lifetimes exhibit considerable scatter about the general log-linear trend. The departures from the exponential fit line are neither random nor normally distributed.   The ensemble of system lifetimes oscillates aperiodically relative to the exponential fit, with typical wavelength and amplitude increasing at larger separations, but remaining  $\lesssim 1$ in $\log t_c$ for the fit interval. The primary cause of this oscillation is the dips in lifetimes associated with mean motion resonances among the planets. However, the most extreme deviations from the exponential fits to system lifetime are the upward spikes in lifetime near $\beta$ = 5.15, 5.41, 5.7 and 6, all of which are located far from the strongest MMRs.  


\subsubsection{ ``Spikes'' of Anomalously Long-Lived Systems}\label{sec:spikes}

Looking more closely at the first spike, 3 systems in the range [5.147, 5.1585] had $t_c>10^6$ years, 2 of which had $t_c>10^7$ years, but no system survived for $>10^8$ years. Given that we performed 2000 runs per unit increment in $\beta$, 2 stable runs corresponds to an effective width of $\sim 0.001$ in $\beta$; i.e., the integral over the phase space of initial conditions of systems with $\beta \approx 5.15$ surviving for more than ten million years is  roughly equal to the value that it would have if all systems within a range in $\beta$ that is 0.001 wide survived  for $>10^7$ years.  Analogously, the lack of any long-lived systems with very tight spacing implies that the effective width of systems with $\beta < 5$ and $t_c>10^5$ years is $\lesssim 0.0005$. 

The numbers of survivors for various time intervals within each of the upward spikes in system lifetimes are listed in Table \ref{tab:spikes}. Systems in the first three spikes for random longitudes were followed for well over an order of magnitude longer than the trend at their separations, and within statistical uncertainties, approximately half of the runs within each of these spike regions went unstable with each increase of one in $\log t_c$.   

Figure \ref{fig:random} displays a wide region of closely-spaced triangles (signifying lower bounds in lifetimes of systems that survived for the entire time interval simulated, $10^8$ years in this case) beginning near $\beta = 6.30$ and ending just above $\beta = 6.70$. We selected an upper boundary for the widest separation spike for random longitudes runs at 6.4095  because the space between it and the next triangle, at 6.444, represented the largest distance between $10^8$ year survivors within that range (and the distance to the subsequent triangle, at 6.4705 is the second largest). This also facilitates comparison of effective widths with the spike having similar boundaries in the SL09 longitudes set of runs.

\begin{table}[htbp]
\caption{ Heights and Widths of Spikes of Long-Lived Closely-Spaced Systems}
\begin{center}
\begin{tabular}{| l | l || c | c | c | c | c |}
\hline
Set of runs &Location & $ t_c >10^6$ yr & $t_c >10^7$~yr & $t_c >10^8$~yr& $ t_c >10^9$~yr& $t_c >10^{10}$~yr\\ \hhline{|=|=||=|=|=|=|=|}
Random & [5.147, 5.1585] & 3; 0.0015 & 2; 0.001 & 0 & 0& 0 \\ \hline 
Random & [5.402, 5.4165] & 7; 0.0035 &4; 0.002 & 2; 0.001 & 1; 0.0005 &  1; 0.0005 \\ \hline
Random & [5.673, 5.773] & -- & 14; 0.007 & 9; 0.0045 & 3; 0.0015& 2; 0.001 \\ \hline 
Random &  [5.9025, 5.9730] &  -- & 9; 0.0045 & 3; 0.0015 &  3; 0.0015 & 3; 0.0015 \\ \hline
Random &  [5.98, 6.0840] &  -- & 18; 0.009 &  13; 0.0065 &  -- &  -- \\ \hline
Random & [6.3055, 6.4095] &  -- &  -- & 37; 0.0185 &  -- &  -- \\ \hhline{|=|=||=|=|=|=|=|}
Primary & [5.377, 5.409] & 18; 0.018 & 13; 0.013 & 7; 0.007  & 2; 0.002 &  1; 0.001  \\ \hline
Primary & [5.92, 6.07] &  -- & 13; 0.13 & 12; 0.12  & 10; 0.1  & 9; 0.09 \\ \hline
Primary & [6.8, 6.9] &  -- & -- & -- &  2; 0.2  & 1; 0.1 \\ \hline
Primary & [7.2, 7.4] &  -- &-- & -- &  3; 0.3 & 2; 0.2 \\ \hhline{|=|=||=|=|=|=|=|}
SL09 &  [5.675, 5.748]  & -- & 30; 0.030 & 8; 0.008 & 2; 0.002 &  0 \\ \hline   
SL09 & [6.33, 6.41] &  -- &  -- & 7; 0.07  & 5; 0.05 & 4; 0.04  \\ \hline
SL09 & [7.15, 7.45] &  -- &  -- & --   & 5; 0.25 & 4; 0.2  \\ \hhline{|=|=||=|=|=|=|=|}
Hexagonal & [5.95, 6.09] &  -- & 11; 0.11 & 10; 0.10  &  --  &  -- \\ \hhline{|=|=||=|=|=|=|=|}
Aligned & [5.131, 5.187] & 18; 0.018 & 7; 0.007 & 5; 0.005  & 4; 0.004  & 2; 0.002 \\ \hline
Aligned & [5.188, 5.255] & 40; 0.040 & 33; 0.033 & 28; 0.028  &  --  &  -- \\ \hline
Aligned & [5.317, 5.358] & 35; 0.035 & 29; 0.029 & 27; 0.027  &  --  &  -- \\ \hline
 Aligned &  [5.87, 6.04]  &  --  &  16; 0.16  & 14; 0.14  &  --  &   -- \\ \hline
\end{tabular}
 \end{center}          
   \tablecomments{Locations (interval from the first system within the spike through that of the last such system), followed by number of runs surviving for the indicated times and effective width in $\beta$ of the ensemble of survivors for that duration.  Spikes are defined by a minimum duration (always chosen to be an integer power of ten) to be considered anomalous, no numbers are quoted for survival times not considered anomalously high for their location, and data are also missing for durations longer than those at which we capped our integrations.   Simulations in the spike for Random longitudes around $\beta = 6$ and that for Aligned systems  around $\beta = 5.2$ were allowed to run for different maximum times depending on the value of orbital separation; in order to represent all results available, the data for each of these spikes are split over two rows. The region in the Primary Set near $\beta = 6$ was  simulated with non-uniform resolution, so only systems with $\beta$ values that are multiples of 0.01 (the lower resolution) are considered here. \label{tab:spikes}  }
 \end{table}

Comparison with the findings of OVT17 for 5-planet systems shows that the most tightly-spaced long-term stable regions occur at smaller $\beta$ and much smaller values of the exponential trend in lifetime for 3-planet systems. Also, the first such regions are  far narrower in $\beta$ and occur in regions where they occupy much smaller fractions of phase space  than the most tightly-spaced stable zones for 5-planet systems.   The first (lowest $\beta$ value) upward spike found for five-planet systems by OVT17 started at $\sim$~8.6, but at that separation the trend in system lifetime is already higher, and the orbital spacing is closer to $\beta$ values where most systems become very long-lived. That spike in lifetimes of 5-planet systems appears to be more closely analogous to the upturns in lifetimes at orbital separations that are ``large'' when expressed in mutual Hill radii that were found by \citet{Duncan:1997} and \citet{Duncan:1998}, who varied the ratio of the masses of the (four or more) secondaries relative to that of the primary rather than the separation between orbits. 

\citet{Obertas:2017} did not find any (five-planet) systems whose lifetimes are much more than one order of magnitude above their exponential fit until $\beta > 8.3$, and by that separation, the typical time to orbit crossing is already several million years. This is the beginning of the approach to their first spike, which is located on both sides of the 13:11 MMR that lies at $\beta \approx 9$.  Note that in said spike, lifetimes probably don't exceed $10^{10}$ years by much, since a large fraction of the systems that survived for 1 Gyr become unstable between $10^{9}$ and $10^{10}$ years. In contrast, typical systems in their spikes at separations of $\beta  \gtrsim 10$  may well persist far longer. Our first spike is at $\beta \approx 5.15$, where the exponential fit gives $t_c < 10^5$ yr; although this spike and the next one near $\beta = 5.41$ are very narrow, a broader spike is present starting around $\beta = 5.7$, where the exponential fit for $t_c$ is still well below $10^6$ yr. 

\subsubsection{ Distribution of Deviations from the General Trend}\label{sec:distdev}

  Table \ref{tab:fit} provides four measures of the dispersion of system lifetime from fits to system lifetime data
 : The standard deviation of $\log t_c$ from the exponential fit is denoted by $\sigma_\mathrm{exp}$. The oscillatory pattern produced by the dips in $t_c$ near resonances is responsible for much of the scatter, so to measure the remaining variations we use the standard deviation from the local rolling median, $\sigma_\mathrm{local}$, where the value of the local rolling mean was computed by averaging the values of $\log t_c$ for the fifth and sixth longest-lived systems among the ten systems with values of $\beta$ nearest to the system being considered (five below and five above). Both of these quantifications of the spread in system lifetimes sum over squares of distances of individual systems from expectations, and thus their values are sensitive to small numbers of outliers, such as anomalously long-lived runs in the spikes; our prescription of stopping integrations at $10^{10}$ years thus also affects these measures.   

\begin{figure}[H]
\centering
\includegraphics[scale=0.75]{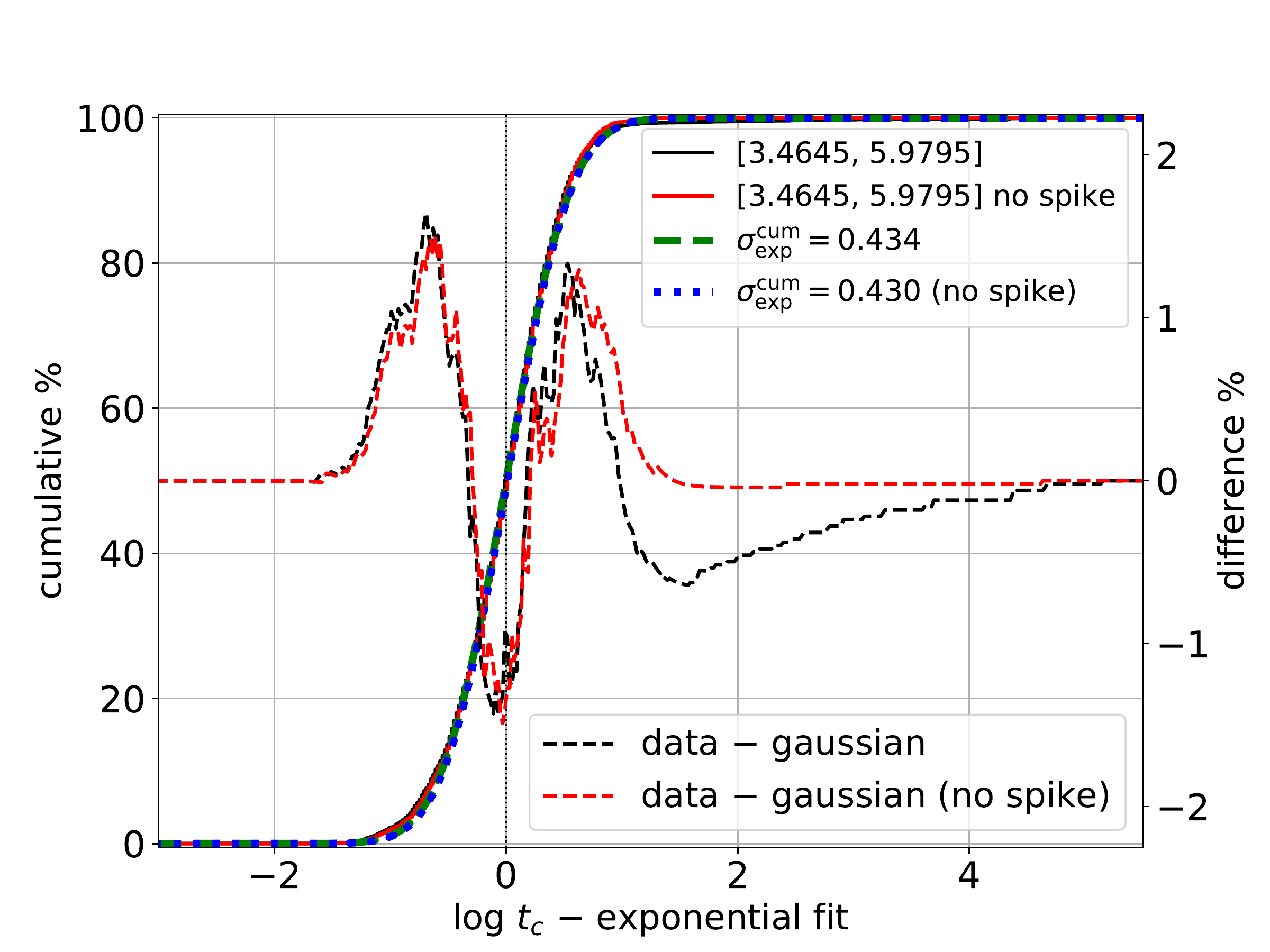}
\caption{Comparison of cumulative distributions of deviations in lifetimes of systems with randomly-drawn initial longitudes in the range $3.4645 \le \beta \le$ 5.9795 from log-linear fits with cumulative best-fit gaussian (i.e., normal) distributions. The four curves associated with the left axis labels are indicated in the key near the top of the figure: The solid black line shows the cumulative portion of the ensemble of runs with lifetimes  ($\log t_c$) less than the amount given by the horizontal coordinate ``above'' the best-fit exponential. The green dashed curve represents a cumulative gaussian distribution with dispersion of 0.434. The solid red and dotted blue curves are analogous representations with the 34 systems with lifetimes $> 3~\sigma_\mathrm{exp}^\mathrm{cum}$ above the exponential fit removed. The two curves associated with the right axis are listed in the lower key: The dashed black (red) curve shows the difference between the solid black (red) ``observed'' cumulative distribution and the dashed green (dotted blue) cumulative gaussian approximation thereto.  The red and blue curves omit exceptionally long-lived runs in the upward spikes of system lifetimes, whereas the black and green curves include all data within the specified range.    \label{fig:deviations_fig2}} 
\end{figure}

Figure \ref{fig:deviations_fig2} shows the cumulative fractional distribution of the ensemble of distances from the $\log t_c$ values for individual systems to the best-fit exponential approximation given in the third numerical row of Table \ref{tab:fit} and various related distributions. The distribution of deviations (solid black curve in Fig.~\ref{fig:deviations_fig2}) is analogous to a  cumulative gaussian centered at zero with standard deviation of $\sigma_\mathrm{exp}^\mathrm{cum} = 0.434$ (dashed green curve). Although the departures from the  cumulative gaussian approximation (dashed black curve in Fig.~\ref{fig:deviations_fig2}) are small, they are non-random. There are 34 systems (0.68\%) with lifetimes that are more than 3 $\sigma_\mathrm{exp}^\mathrm{cum}$ above the exponential fit (compared to 0.13\% for a gaussian distribution) and 16 systems (0.32\%) exceed 6 $\sigma_\mathrm{exp}^\mathrm{cum}$ (compared to $1\times 10^{-7}$\% for a gaussian). All of these exceptionally long-lived systems have separations that place them within the upward spike regions near $\beta = 5.15, 5.41$ and 5.7 or the first portion of the next spike that begins near the upper end of the range in $\beta$ considered here. If one excludes these 34 anomalously long-lived systems,  the slopes flattens slightly to $b' =1.3892$, the ``intercept'' goes up a little to $c' = 2.1005$,  the standard deviation of the points to the exponential fit $\sigma_\mathrm{exp}$ drops by 12\% to 0.438, but the best fitting cumulative gaussian dispersion value  drops by just 1\%, to $\sigma_\mathrm{exp}^\mathrm{cum} = 0.430$. We include values $\sigma_\mathrm{exp}^\mathrm{cum}$ and the analogous $\sigma_\mathrm{local}^\mathrm{cum}$ in Table \ref{tab:fit} because, unlike the standard measures of dispersion, their values are not sensitive  to small numbers of points far from the fit curve. 
The ratio $\sigma_\mathrm{exp}^\mathrm{cum}/\sigma_\mathrm{local}^\mathrm{cum}$ is $\sim 1.6$ for all three intervals in $\beta$ examined for the set of runs with Random longitudes; as deviations add in quadratures and this ratio  exceeds $\sqrt 2$,  the variance of the local trend relative to the exponential fit must be larger than the variance of individual points relative to the local trend. 

\begin{figure}[H]
\centering
\includegraphics[scale=0.75]{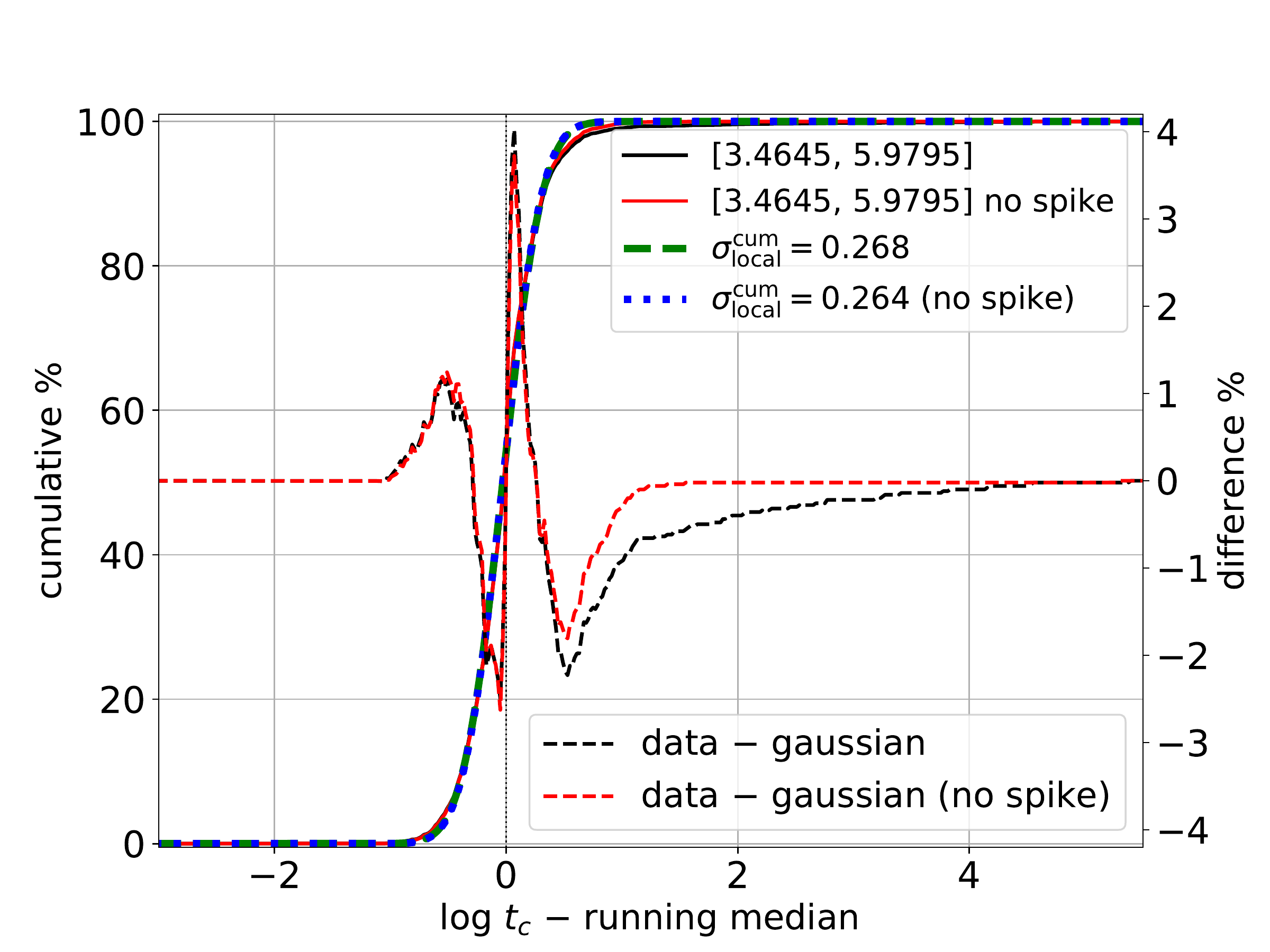}
\caption{{ Comparison of cumulative distribution of deviations in lifetimes of systems with randomly-drawn initial longitudes in the range $3.4645 \le \beta \le$ 5.9795 from log-linear fits with local median. The four curves associated with the left axis labels are indicated in the key near the top of the figure: The solid black line shows the cumulative portion of the ensemble of runs with lifetimes  ($\log t_c$) less than the amount given by the horizontal coordinate ``above'' the local median. The green dashed curve represents a cumulative gaussian distribution with dispersion of 0.268. The solid red and dotted blue curves are analogous representations with the 34 systems with lifetimes $>3~\sigma_\mathrm{exp}^\mathrm{cum}$ above the exponential fit removed. The two curves associated with the right axis are listed in the lower key: The dashed black (red) curve shows the difference between the solid black (red) ``observed'' cumulative distribution and the dashed green (dotted blue) cumulative gaussian approximation thereto.  The red and blue curves omit exceptionally long-lived runs in the upward spikes of system lifetimes, whereas the black and green curves include all data within the specified range.} \label{fig:deviations_median}} 
\end{figure}

The red curves in Fig.~\ref{fig:deviations_fig2} show the  cumulative distribution excluding the 34 points in the spikes that survive for more than  3~$\sigma_\mathrm{exp}^\mathrm{cum}$ longer than the prediction from the exponential fit.  The red dashed curve showing the difference between this   cumulative distribution and the best-fit gaussian never exceeds $\sim 1.5$\% of the \emph{total number of runs}.  Nonetheless this difference between the two distributions  is  statistically significant. For example, 
there are 95 points in the interval (--2, --1), but only 29 points in the interval (1, 2), whereas the gaussian approximation predicts the same number of points within these regions,  a difference that greatly exceeds expected stochastic variations. The dashed red difference curve is largely symmetric about zero.  But this isn't the type of symmetry that would be present if the distributions of deviations were the same on the upside as on the downside - that would be antisymmetric, with values above the mean being equal to negative those for the same amount below the mean.  These curves show that the narrow resonant dips cause  there to be more systems with lifetimes that are roughly 2$\sigma_\mathrm{exp}^\mathrm{cum}$ below the fit line than expected  (producing the rise to the positive peak near --0.6 on the abscissa and bringing the fit line below the majority of points), so there are fewer systems than expected below the fit line by $\lesssim 1.5\sigma_\mathrm{exp}^\mathrm{cum}$ (thus causing the drop to the most negative value near 0 on the abscissa),  more than expected above the fit line by $\lesssim 1.5\sigma_\mathrm{exp}^\mathrm{cum}$, and a deficit of systems (outside of the spikes) that survive $\gtrsim 1.5\sigma_\mathrm{exp}^\mathrm{cum}$ above the fit line. The larger deviations of the points below the line is compensated for by the median being above the mean.  This is because, excluding the upward spikes, the largest deviations from the fits are the (downward) dips near the strongest resonances.

 Much of the deviations of lifetimes from the best fitting exponential can be attributed to the oscillatory pattern caused by dips in lifetimes near strong resonances, although there is also some scatter about the local average. Figure \ref{fig:deviations_median} is analogous to Figure \ref{fig:deviations_fig2} (same color code and same range), but shows the cumulative fractional distribution of the ensemble of distances between $\log t_c$ of individual systems and the local running median defined above as the average of  $\log t_c$ of the fifth and sixth longest-lived systems among the ten systems with most similar spacing.  The solid black curve is the distribution of the differences, which has a standard deviation of $\sigma_\mathrm{local}^\mathrm{cum} = 0.268$, much smaller than the standard deviation of $\sigma_\mathrm{exp}^\mathrm{cum} = 0.434$ found by comparing with the best-fit exponential. The dashed red curve in  Fig. \ref{fig:deviations_median}, representing the differences of the cumulative distribution from a gaussian with the same standard deviation, is nearly antisymmetric about the vertical axis, implying a symmetric distribution of deviations from the local mean.  The wings are higher than for a gaussian, compensated for by an excess of systems with lifetimes very similar to the running median.

We performed a similar analysis of the crossing times of five-planet systems integrated by \citet{Obertas:2017}.  Results for four different upper limits to the range of orbital separations analyzed are listed in Table \ref{tab:fit_ovt17}. Five-planet systems have smaller deviations from the log-linear fit over the same range in orbital separation that we examined for three-planet systems with randomly-selected longitudes (Table \ref{tab:fit}), but the deviations are substantially larger when the entire zone prior to the first regions of long-term stability is considered. The increases from the smallest region to the one extending to $\beta = 8.3$ are primarily caused by the increasing depth of the resonant dips, whereas those for the two largest regions fit are dominated by the first upward spike in the OVT17 data. Regardless of the range examined,  5-planet systems have characteristic variations from the local median lifetime, $\sigma_\mathrm{local}^\mathrm{cum}$, that are $\sim 25\%$ smaller than those seen with 3-planet systems with Randomly chosen longitudes.

  \begin{table}[htbp]
\caption{ Coefficients of Log-Linear Fits to  Lifetimes of Five-Planet Systems}
\begin{center}
\footnotesize
\begin{tabular}{| l | l | l || c | c | c | c | c | c | c | c |}
\hline
Set of runs & Range &  \multicolumn{1}{l ||}{Resol'n} & \multicolumn{1}{c|}{ Slope, $b'$} & \multicolumn{1}{c|}{Intercept, $c'$} & \multicolumn{1}{c|}{$\beta$ at $\log t_c = 10$} & \multicolumn{1}{c|}{$t_c >10^{10}$} &{\textbf{$\sigma_\mathrm{exp}$}} &  \multicolumn{1}{c|}{\textbf{$\sigma_\mathrm{local}$}} &   \multicolumn{1}{c|}{\textbf{$\sigma_\mathrm{exp}^\mathrm{cum}$}} & \multicolumn{1}{c|}{\textbf{$\sigma_\mathrm{local}^\mathrm{cum}$}}    \bigstrut \\ \hline

 OVT17 &  $2\sqrt{3}$ -- 5.9795  & 0.0005  & 1.082 $\pm$ 0.007 & 1.940 $\pm$ 0.011 & 10.914 $\pm$  0.041 & 0 & 0.378 &  0.216 &  0.379 &  0.201 \\ \hline 

OVT17  &  $2 \sqrt{3}$ -- 8.3 &  0.0005 &  0.936 $\pm$ 0.004 &  2.121 $\pm$ 0.011 &  11.884 $\pm$ 0.024 & 0 & 0.525 &  0.213 &  0.519  &  0.194\\ \hline

 OVT17  &  $2 \sqrt{3}$ -- 8.7 & 0.0005 &  1.044 $\pm$ 0.004 &  1.934  $\pm$ 0.012 &  11.184 $\pm$ 0.019 &  9 &  0.617 &  0.223 &  0.534  & 0.198\\ \hline

 OVT17 & $2 \sqrt{3}$ -- 9.2 &  0.0005 &  1.168 $\pm$ 0.004 & 1.714 $\pm$ 0.014 &  10.556 $\pm$ 0.029 & 196 &  0.734 &  0.242 &  0.639 &  0.205 \\ \hline
\end{tabular}
 \end{center}          
   \tablecomments{{ Values of the  coefficients computed from the exponential fits to the lifetimes of more than ten thousand systems integrated by OVT17. From left to right, the columns in this table list the name of the Set, range  of initial orbital separation, resolution in $\beta$, the derived coefficients $b'$ and $c'$, the value of $\beta$ where the orbital crossing time predicted by the fit line is $10^{10}$ years,  the number of runs that survived for the entire $10^{10}$ years simulated, and the four measures of dispersions of the data listed for three-planet systems in Table \ref{tab:fit}}.  \label{tab:fit_ovt17} }
 \end{table}  

\medskip

\section{Initial Longitudes Approaching Conjunction}\label{sec:results}

Our Primary Set of integrations\footnote{We initiated our study using these initial planetary longitudes and we integrated most of the systems in this set prior to realizing how much lifetimes at a given orbital separation could depend on initial longitudes. Once we became aware of these dependances, we decided to consider other starting longitudes, but allowed fewer of those integrations to proceed beyond $10^8$ years because of computational costs.} 
consisted of three-planet systems with the same set of initial longitudes of each of the planets used for all integrations in the Set, as described in \S\ref{sec:methods} and enumerated in the second row of Table \ref{tab:initial}.  Each system began with a different value of orbital separation $\beta$, and every system was followed until a pair of orbits crossed or the simulation time reached a pre-determined limit of either $10^8$ years or $10^{10}$ years.  System lifetimes ranged from $< 100$ years to $> 10^{10}$ years, with the longest-lived systems being computationally expensive.  Thus, to maximize scientific returns from our computational resources, we explored parameter space with nonuniform resolution in $\beta$. We covered the region $3.465 \leq \beta \leq 6.750$ using steps of 0.001 in $\beta$ in most of this interval, but only considering multiples of  0.01 in $\beta$ within the long-lived regions $5.91 <\beta<5.99$,  $6.016 <\beta<6.08$ and  $6.62 <\beta$. 
For $\beta> 6.75$, a large fraction of the systems were very long-lived, 
so we stopped most systems that had not already become unstable at $10^8$ years, although we followed simulations for  all multiples of 0.1 from $\beta =6.8$ to $\beta = 7.5$, as well as for the separations $\beta =8$ and $\beta = 8.5$, until instability or  $10^{10}$ years had elapsed. 

\begin{figure}[H]
\centering
\includegraphics[scale=0.75]{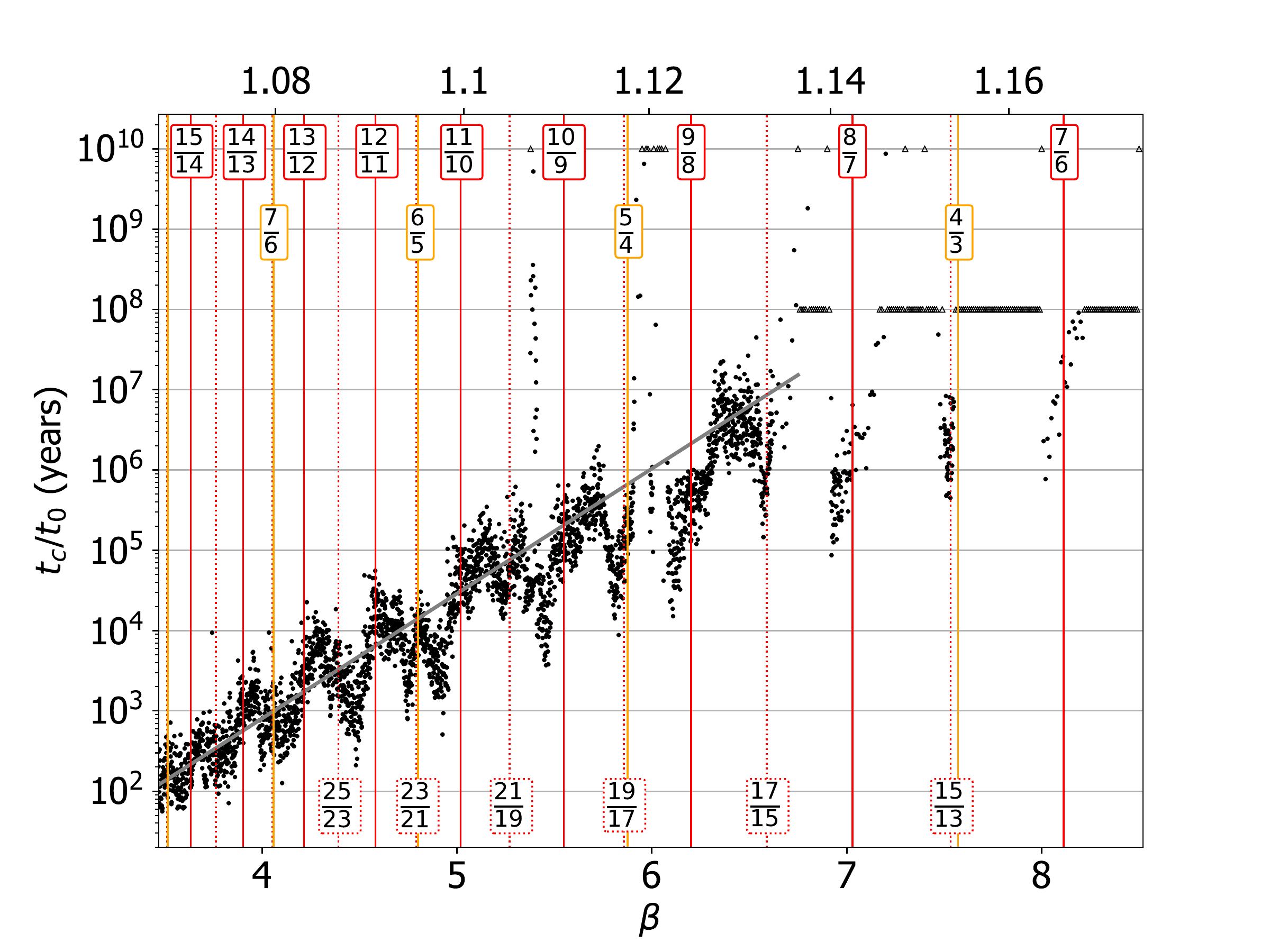}
\caption{Lifetime, $t_c$, of three-planet systems as a function of the initial separation of the orbits of neighboring planets in units of mutual Hill sphere radii, $\beta$ (Eq.~\ref{eq:beta}), for our Primary set of runs. The  open triangles represent systems that remained stable for the entire time interval, either $10^8$ or   $10^{10}$ years, simulated. The solid red vertical lines denote the locations of first-order mean motion resonances between adjacent pairs of planets, the  dotted red vertical lines denote the locations of second-order MMRs between adjacent pairs of planets, and the solid gold vertical lines denote first-order resonances between the innermost and outermost planets. The gray line represents the log-linear fit to all points within the range $3.47 \le \beta \le 6.75$ at 0.01 resolution.} 
\label{fig:uniform}
\end{figure}

 Figure \ref{fig:uniform} shows the crossing time, $t_c$, as a function of initial orbital separations given in mutual Hill radii and as a function of initial period ratios; all runs from our Primary set of integrations are represented. The overall trend is again that more widely spaced systems tend to be more stable, but there remains considerable variation in system lifetimes about this general trend. We expect that mean motion resonances (MMRs) and chaos contribute to this scatter, and we analyze these effects in \S\ref{sec:resonances} and \S\ref{sec:chaos}, respectively.

We fit our data at multiples of 0.01 in $\beta$ over the interval $3.47 \leq \beta \leq 6.75$ to  the form of Equation (\ref{eq:exponential}) and computed values of the constants to be $b = 1.5559$ and $c = -3.3186$. Extrapolation of this exponential fit to more widely-spaced systems estimates the separation at which $t_c = 10^{10}$ years to be $\beta=8.5602$. Globally, this fit is consistent with the results of SL09, who found $b = 1.496$ and $c = -3.142$ (their fit covered the range $3 \le \beta \le 7.2$ at 0.1 resolution, plus points at 7.5 and 8). As with the crossing times of runs with random initial longitudes, the data show significant scatter about the general trend. To isolate the contribution of the initial longitudes to this scatter,  in \S\ref{sec:deviations} we compare this scatter to that found in  \S\ref{sec:deviationsOVT}.

\subsection{Deviations from the Exponential Fits}\label{sec:deviations}
   
 The cumulative measures of dispersion of lifetimes for Primary runs are smaller than those of Random longitudes runs computed over the same interval in orbital separation, with values of $\sigma_\mathrm{exp}^\mathrm{cum}$ lower by $\sim 5\%$,  and those for $\sigma_\mathrm{local}^\mathrm{cum}$ are down by $\sim 13\%$ (Table \ref{tab:fit}). These differences provide a measure of the increase in dispersion of lifetimes when varying initial starting angles in addition to orbital separations. Note that no systematic differences are present when comparing standard deviations of lifetimes from the exponential and local fits, $\sigma_\mathrm{exp}$ and $\sigma_\mathrm{local}$,  because those estimates are much more sensitive to the numbers of systems in the long-lived spikes. 

Table  \ref{tab:fit} also provides fit parameters for systems with larger orbital separation, where the resolution in  $\beta$ is 0.01.     The local median can change nontrivially over distances of 0.05 in $\beta$, leading to systematic increases in  $\sigma_\mathrm{local}$ for fits to data at resolution of 0.01.  Thus, we caution the reader against overinterpreting these statistical measures. We don't include fits for separations at which only 0.1 resolution data are available up to $10^{10}$ years because statistics are poor for both measures and  $\sigma_\mathrm{local}$ extends well beyond the local region.
 
As was the case for systems with randomly-selected initial planetary longitudes, the departures from the exponential trend in system lifetimes are neither random nor normally distributed. As for the set of systems with random, independently drawn, initial longitudes, the ensemble of system lifetimes oscillates aperiodically relative to the exponential fit,  and the most extreme deviations from the exponential fits  to system lifetime occur in upward spikes.  The key difference is that there are fewer upward spikes for this fixed set of initial longitudes, but those that are present have more members (larger effective width).  In the region [3.47, 6.75] at 0.01 resolution in $\beta$, the median difference in absolute value of $\log t_c$ of neighboring  points (lifetimes of systems having the closest values of $\beta$, excluding pairs of points that were both stopped after $10^{10}$ years)  is 0.442. The signs of the differences in lifetime of neighboring systems from the general trend line are positively correlated, with 68\% of runs having $t_c$  on the same side of the log-linear fit line as the run with $\beta$ value 0.01 smaller (compared with 50\% if no correlation existed).   

\begin{figure}[H]
\centering
\includegraphics[scale=0.75]{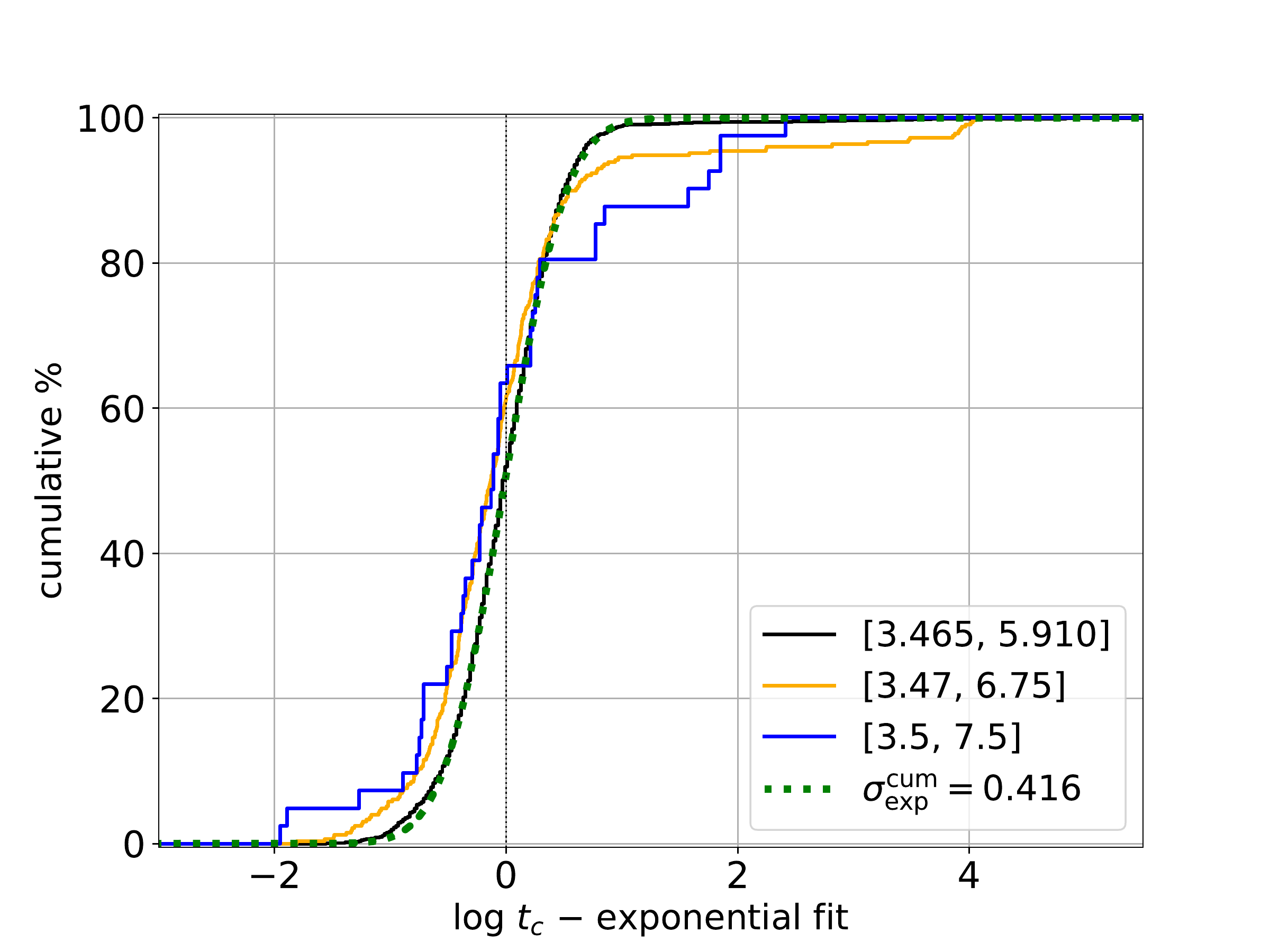}
\caption{Cumulative portion of the ensemble of Primary longitudes runs with lifetimes  ($\log t_c$) less than the amount given by the horizontal coordinate ``above'' the best-fit exponential. Three different ranges and resolutions in $\beta$ are shown by solid lines; specifics are presented in the color key in the lower right. The green dotted curve represents a cumulative gaussian (i.e., normal) distribution with dispersion of $\sigma_\mathrm{exp}^\mathrm{cum} = 0.416$.\label{fig:deviations}} 
\end{figure}

    Figure \ref{fig:deviations} shows the cumulative fractional distribution of the ensemble of distances between $\log t_c$ and the best-fit exponential approximations given in the fifth, seventh and eighth numerical rows of Table \ref{tab:fit}. The distribution up to 5.910 (black curve in Fig.~\ref{fig:deviations}) is analogous to a cumulative gaussian distribution centered at zero with standard deviation $\sigma_\mathrm{exp}^\mathrm{cum} = 0.416$ (dotted green curve), only $\sim 5$\% smaller than the dispersion over a similar range in planetary separation for systems with randomly-drawn initial longitudes. As for systems with randomly-drawn longitudes, the gaussian fit is a fairly good approximation apart from the excess of unusually long lived systems: 0.82\% (20 systems) have lifetimes that are $> 3~\sigma_\mathrm{exp}^\mathrm{cum}$  above the exponential fit (compared to 0.13\% for a gaussian distribution) and 0.45\% (11 systems) exceed 6 times the standard deviation (compared to $1\times 10^{-7}$\% for a gaussian). Most of these exceptionally long-lived systems have separations that place them within the upward spike region near $\beta = 5.38$; the exceptions are the system with $\beta = 3.744$, which has a lifetime $3.4\sigma_\mathrm{exp}^\mathrm{cum}$ above the fit line (the system with $\beta = 3.691$ has a lifetime $2.9\sigma_\mathrm{exp}^\mathrm{cum}$ above the fit line), and the systems with the two largest values of $\beta$ within this interval, 5.909 and 5.91, which are $3.8\sigma_\mathrm{exp}^\mathrm{cum}$ and $3.1\sigma_\mathrm{exp}^\mathrm{cum}$ above the fit line, respectively. The other curves deviate more visibly from normal distributions, because the percentage of their points within  spikes is much larger, leading to a displacement of the median (where the curves on the figure reach 50\%) from 0 to balance the significant depressions of the curves from the normal shape at positive values above the standard deviation. These departures from normality would be even larger if we had the computer resources to integrate those systems that survived for $10^{10}$ years for even longer times.   
 
The spikes in system lifetimes seen for randomly-selected initial longitudes near $\beta$ = 5.15, 5.7 and 6.35 are absent in Fig.~\ref{fig:uniform}, but the very narrow spike centered at $\beta \approx 5.385$,  the wider long-lived region around $\beta = 6$, and the even wider stable region starting at $\beta = 6.75$ remain.  Looking more closely at the first spike, 18 systems in the region [5.377, 5.409] had $t_c>10^6$ years, 13 of which had $t_c>10^7$ years, including 7 that had $t_c>10^8$ years, 2 with $t_c>10^9$ years, and 1 with $t_c>10^{10}$ years. As with the spikes seen in the Random longitude runs with $\beta < 5.8$, within statistical uncertainties, approximately half of the runs within the spike region went unstable with each unit increment in $\log t_c$. The numbers of surviving systems imply an effective width in $\beta$ of $\approx 0.018$ for survival time of $10^6$ years, dropping to $\sim 0.001$ for survival time of $10^{10}$ years. However,  lifetimes are distributed differently within the second spike, with 9 of the 13 runs at multiples of 0.01 in $\beta$ that survived for more than 10 Myr surviving for the full 10 Gyr integrations (Table \ref{tab:spikes}). Both types of dropoff are present in some of the spikes detected at other initial longitudes; see Section \ref{sec:discussion} for details.

\section{Resonances and displacement of resonant dips}\label{sec:resonances}

SL09 identified a drop in the lifetime of systems associated with a first-order mean motion resonance (where the ratio of the orbital periods of a pair of planets can be written in the form $N/(N-1)$, with $N$ being an integer) between neighboring planets.  First-order resonances of neighboring planets have the strongest effects on lifetimes.  They consist of resonance triplets -- both the inner and outer pairs are near resonance (these resonances are slightly offset, by a fractional difference of order M$_\oplus$/M$_\odot$, but this offset is small compared to the resonance widths), and the planets are near a (zeroth-order) 3-body resonance of the form $Nn_j -(2N+1)n_{j+1}+(N+1)n_{j+2} =0$, where $n_j$ is the angular velocity of the $j^{\rm th}$ planet. 

\citet{Obertas:2017} showed dips in lifetime at several first-order and second-order (period ratio  $N/(N-2)$) resonances, and they discussed resonances at greater length. They  found that the dips near first-order MMRs were centered at slightly smaller values of $\beta$ than the resonance locations, but those at second-order MMRs were centered at the resonance locations. However, they were unable to explain the cause of the displacements for first-order resonances. Our data show that the locations of the dips in lifetimes at second-order resonances are displaced inwards, but by far smaller amounts than the displacement of those produced by first-order resonances (Fig.~\ref{fig:uniform}). 

\begin{figure}[H]
\centering
\includegraphics[scale=0.75]{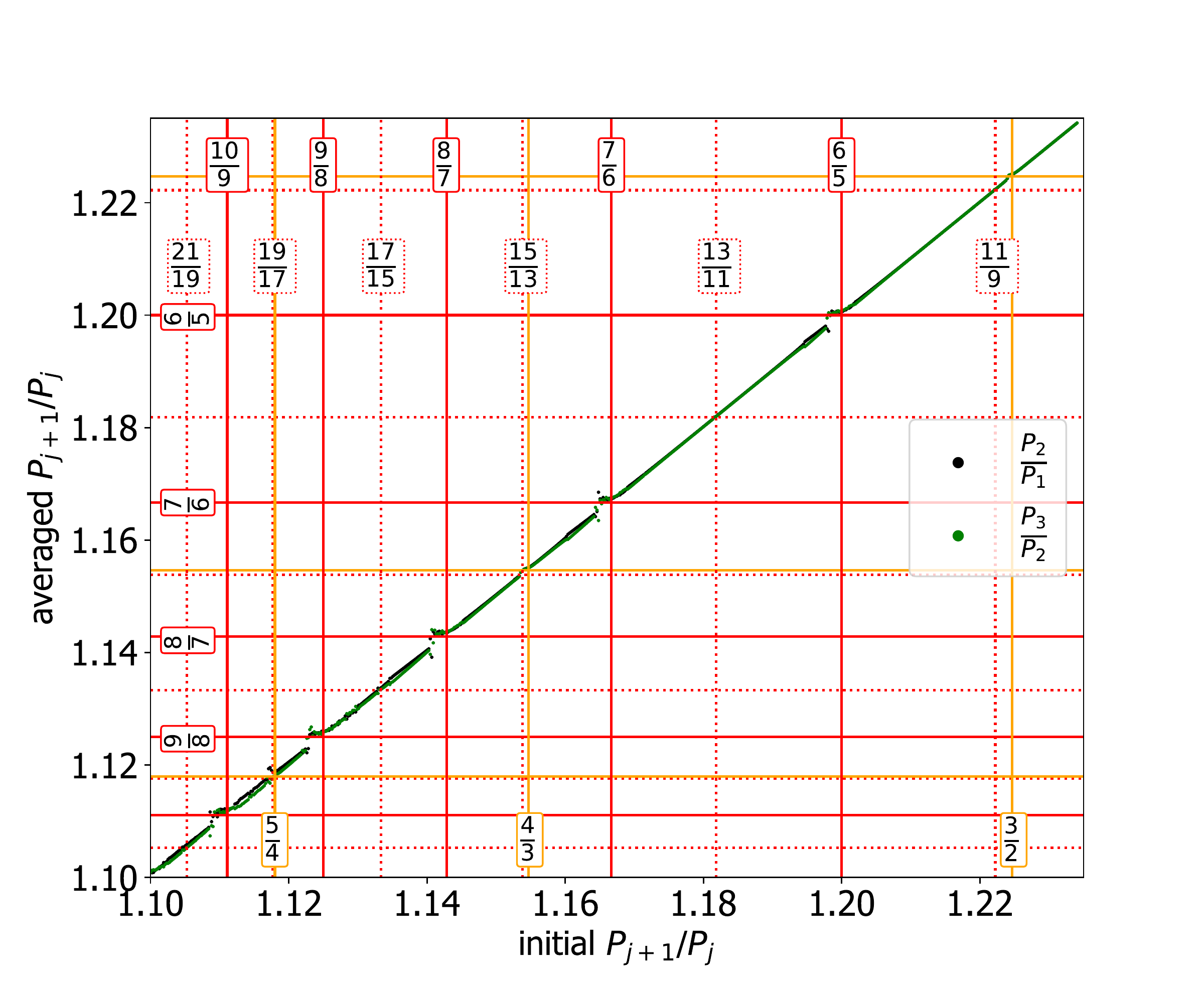}
\caption{Mean orbital period ratios of adjacent pairs of planets as a function of their initial period ratios. The horizontal axis is the initial period ratio of neighboring planets. The vertical axis shows the period ratios of the inner pair of planets (black dots) and the outer pair (green dots) of planets averaged over the first 10,000 years of evolution (1000 years for initial period ratios $< 1.12$, because some of those runs survived $\lesssim 10^4$ years). The solid red vertical (horizontal) lines show the locations of the first-order MMRs between the initial (averaged) orbital periods of neighboring planets, the dotted red lines show the second-order MMRs between neighboring planets and the solid gold lines are first-order MMRs of the inner and outer planets. The runs were spaced by  0.01 in $\beta$.  Note that the left and lower boundaries of the box correspond to the locations of the 11:10 resonance between neighboring planets.} 
\label{fig:average}
\end{figure}

To investigate the shift of the location of the resonant dips, we measured the average orbital period ratios over the first 10,000 orbits (the first 1000 orbits for the most closely-spaced systems) and compared them to the initial ratios calculated using Kepler's Third Law, modified by the mass(es) of the inner planet(s) where appropriate. Planets began at the same longitudes used for the integrations presented in \S\ref{sec:results}. We also consider resonances between the inner planet and the outer planet. Figure \ref{fig:average} plots the averaged period ratios as functions of the initial period ratios. As the dynamical separations  between the inner pair and the outer pair of planets are equal, the initial period ratio is the same for both pairs of adjacent planets. We find that averaged orbital period ratios significantly exceed initial period ratios near and especially immediately interior to the nominal locations of first-order MMRs of adjacent planets.  Many systems have average period ratios very close to the resonant ratios  because these systems initially reside within the resonant island or its chaotic layer. 
Three of the six second-order MMRs between adjacent planets in the region shown  are located near first-order resonances between the inner and the outer planets, and these regions in $\beta$ exhibit analogous, albeit smaller, patterns of increases in average period ratio.  In contrast, the `isolated' 21/19, 17/15 and 13/11 second-order MMRs of neighboring planets had minimal increases in mean orbital period ratios (we find a slight inward displacement of the lifetime dips caused by these resonances, i.e., they are centered very near but not at the second-order MMRs of neighboring planets).  

The period ratios of planet pairs near first-order MMRs  increase as eccentricities are excited by the resonances (see, e.g., \citealt{Lissauer:1998}), whereas the strength of second-order MMRs are proportional to planetary eccentricities, which are initially zero in all of our simulations. The ratio of energy to angular momentum needed to expand a circular orbit of the $j^{th}$ planet is equal to said planet's mean motion, $n_j$.  Thus, excitation of eccentricities of one or both members of a pair of interacting planets requires the planetary orbits to spread apart \citep{Lissauer:1985}. The process involved is analogous to that which requires viscous accretion disks, including planetary rings, to spread radially \citep{Goldreich:1978, Lynden-Bell:1974}.

Although planetary orbits can be stabilized by resonant locks, models of the TRAPPIST-1 planetary system, which contains a resonance chain of seven planets \citep{Gillon:2017, Luger:2017, Tamayo:2017}, have shown that even for far more widely-spaced systems than those considered herein the stable resonant regions in high multiplicity planetary systems occupy a small volume of parameter space. This smallness is consistent with the apparent absence of systems stabilized (for much longer than non-resonant systems with similar separations) by resonant locks among the many thousands of systems that we (and OVT17) simulated.
Libration of planets within resonances could nonetheless explain individual systems that were extremely long-lived compared to most systems with similar values of $\beta$, but the clustering of these exceptionally stable systems with quite similar values of $\beta$, their locations far from the strongest resonances, and the similarity of these stable regions apart from increases in width for higher $\beta$, strongly suggest that the primary reason for stability is being \emph{distant} from major resonances.  

\section{SL09 Longitudes}\label{sec:SL09} 

Following SL09, we started the planets in this Set of simulations at widely-separated initial longitudes, as described in \S\ref{sec:methods} and enumerated in the third numerical row of Table \ref{tab:initial}. Our study covered the range 3.465 -- 6.300 with resolution of 0.001 in $\beta$ and 6.31 -- 8.5, as well as the near-resonance regions 8.75 -- 8.82,  9.56 -- 9.63 and 10.52 -- 10.59,
 with resolution of 0.01 in $\beta$. Our integrations were allowed to continue up to 10$^{10}$ years for $\beta \le 7.1$5. In order to reduce computational costs, most integrations for wider separations (larger values of $\beta$) were halted at $10^8$ years if no orbits had crossed; however, simulations were followed for as long as 10$^{10}$ years for $\beta$ values that are multiples of 0.05 up to $\beta = 7.5$. No SL09 longitudes run with $\beta > 8.18$ went unstable during the $10^8$ years integrated.

 \begin{figure}[H]
\centering
\includegraphics[scale=0.75]{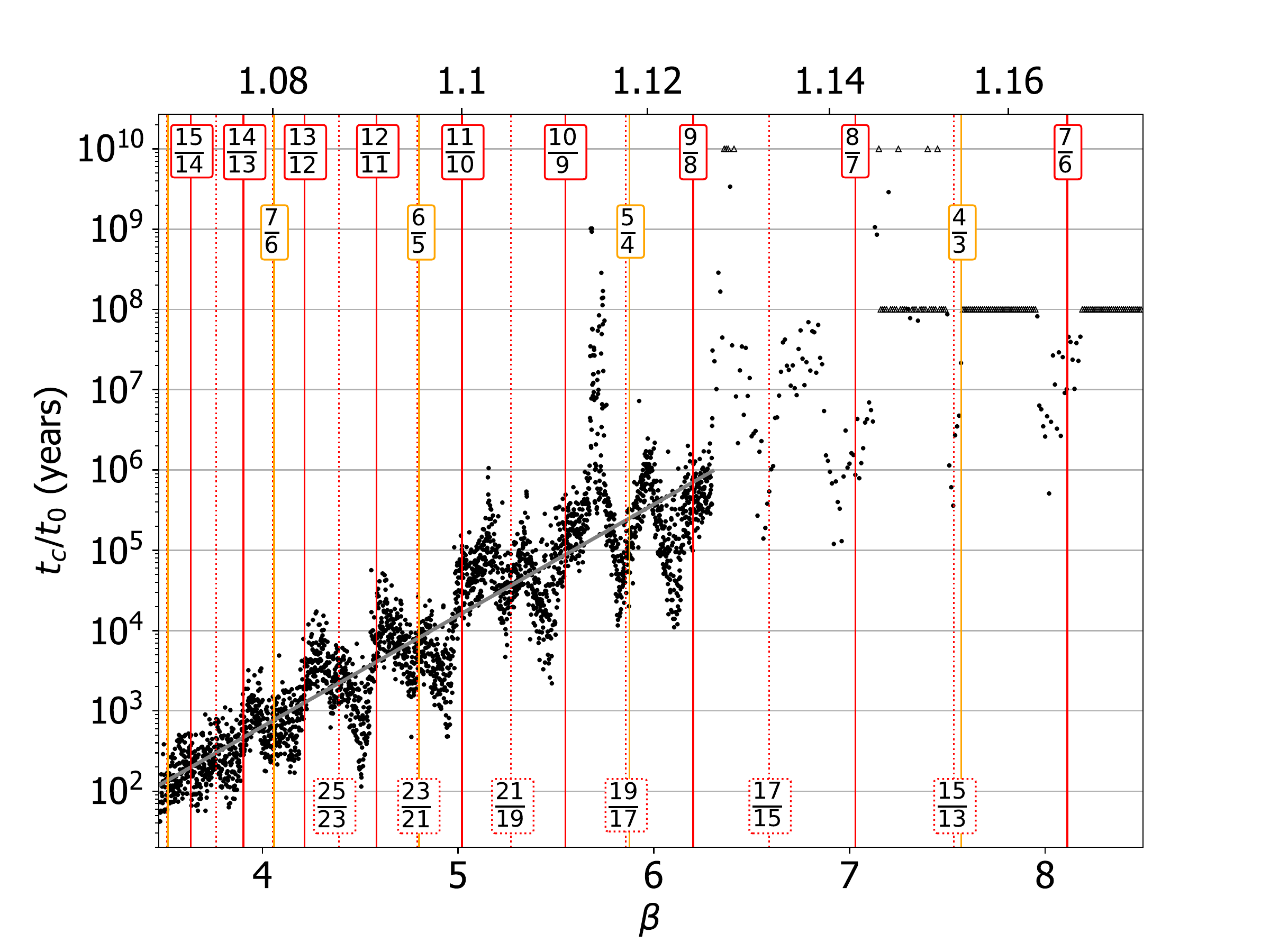}
\caption{Lifetime, $t_c$, of three-planet systems with the same initial longitudes  as those used by SL09 are displayed as a function of the separation in units of Hill sphere radii, $\beta$ (Eq.~\ref{eq:beta}). We used a resolution of 0.001 in $\beta$ for the range $[3.465, 6.300]$ and followed all of these runs until a pair of orbits crossed. For larger separations, in the range $[6.31, 8.50]$, we chose a coarser resolution of 0.01 in $\beta$; we followed systems with $\beta \le 7.1$5 and those with $\beta$ values that are multiples of 0.05 and $\le 7.5$ for up to $10^{10}$ yrs, but stopped other runs if they reached  $10^{8}$ yrs. The  open triangles represent systems that remained stable for the entire $10^{8}$ or $10^{10}$ year time interval simulated. The gray line corresponds to the fit over the range $3.465  \le \beta \le 6.300$ at 0.001  resolution, $\log{t_c} = 1.3834(\beta  - 2\sqrt{3}) + 2.0635$ (Table \ref{tab:fit}).  } 
\label{fig:SL09}
\end{figure}

Figure \ref{fig:SL09} shows that, as for systems starting at our Primary Set of longitudes (Fig.~\ref{fig:uniform}), some tightly-spaced systems far from first-order MMRs survive for very long times. For the SL09 longitudes, the ``anomalous peaks'' (spikes) in system lifetimes are seen near $\beta \approx 5.7$, 6.4 and  7.3. 
 (The most closely-spaced three-planet system that SL09, who used far lower resolution in $\beta$, found to be long-lived began with a separation of $\beta = 6.4$.) In contrast to the lowest $\beta$ peaks for our Primary runs, all of these  spikes are located midway between a first-order resonance between neighboring planets on the inside and a second-order resonance at higher $\beta$. No analogous long-lived region exterior to a first-order resonance between neighboring planets and interior to the neighboring second-order resonance is present for $\beta < 7.5$.

We find eight systems with $\beta < 6.300$ that survived more than $10^{8}$ yrs. Among these eight systems, two survived more than $10^{9}$ yrs: one with initial $\beta = 5.678$ that survived 1.016 Gyr and another with initial $\beta = 5.684$ that survived 1.021 Gyr. (The system with initial $\beta = 5.683$  survived for 931 Myr.)
 
Weak peaks in system lifetime are present near the planetary separations ($\beta$ values) of the two most tightly-spaced spikes  in lifetimes seen for our random initial longitudes runs (Fig.~\ref{fig:random}), with the second peak being substantially less prominent at this combination of longitudes than for random longitudes. The effective width of the  spike located near $\beta = 5.7$ is 4.3 times as large as that computed for random orientations for lifetimes exceeding $10^7$ years, but survival rates of the remaining systems are lower for later times, leading to similar equivalent widths on gigayear time scales (Table \ref{tab:spikes}).

\begin{table}[htbp] 
\caption{ Comparison of System Lifetimes: Primary vs.~SL09 Longitudes}
\scriptsize
\begin{center}
\begin{tabular}{| l  r || c | c | c || c ||| c | }
\hline
Interval in $\beta$ && \multicolumn{1}{c|}{[3.465, 3.999]} & \multicolumn{1}{c|}{[4.000, 4.999]} & \multicolumn{1}{c||}{[5.000, 5.910]} & \multicolumn{1}{c|||}{[3.465, 5.910]} & \multicolumn{1}{c|}{[5.92, 6.75]} \bigstrut \\ \hline
number of runs in the range &  & 535 & 1000 & 911 & 2446 &  84 \\ \hline
$<\log{t_c}$(Primary)$-\log{t_c}$(SL09)$>$ &  & 0.134 & 0.130 & $-$ 0.003 & 0.081  & 0.408   \\ \hline
$<|\log{t_c}$(Primary)$-\log{t_c}$(SL09)$|>$ &  & 0.313 & 0.337 & 0.470 & 0.381 & 1.289    \\ \hline

\# ${t_c}$(SL09) $<0.5{t_c}$(Primary) &  & 169 (31.59\%) & 326 (32.6\%) & 216   (23.7\%) & 711  (29.1\%) &  33  (39.3\%) \\ \hline
\# $0.5{t_c}$(Primary)$<{t_c}$(SL09) $<2{t_c}$(Primary) &  & 293 (54.8\%) & 533 (53.3\%) & 475 (52.1\%) & 1301 (53.2\%) &  27 (32.1\%) \\ \hline
\# $2{t_c}$(Primary)$<{t_c}$(SL09) &  & 73 (13.65\%) & 141 (14.1\%) & 220 (24.1\%) & 434 (17.7\%) &  24  (28.6\%) \\ \hline

\# within 10\% of the Primary runs &  & 40 (7.48\%) & 71 (7.10\%) & 60  (6.59\%) & 171 (6.99\%) &  8  (9.52\%) \\ \hline
\# within 1\% of the Primary runs &  & 3 (0.56\%)*  &  7 (0.70\%)* & 2 (0.22\%)* & 12 (0.49\%)*&  0 (0.00\%)\\ \hline
 Kolmogorov-Smirnov test (p-value) & & 6.12 $\times$ 10$^{-5}$  & 4.54 $\times$ 10$^{-7}$ & 0.099  & 0.002 & 0.175  \\
\hline
\end{tabular} 
\end{center}
\tablecomments{ \label{tab:primary}Comparison of the survival times of integrations using the SL09 initial longitudes runs with those at the same value of $\beta$ from our Primary integrations. Note that because we check for orbital crossing once per year for $\beta < 5$ and once every ten years for more widely-spaced systems, the listed data are unreliable for runs with short lifetimes. Thus,  the data denoted with $^*$ are likely to be somewhat in error, but are nonetheless useful for comparison with the values presented in Tables \ref{tab:chaos} -- \ref{tab:aligned}. Small errors may be present elsewhere in the table, but these are likely to be far less than the statistical uncertainties resulting from the small sizes of the samples.}
\end{table}

 We perform statistical studies by comparing lifetime of the SL09 longitudes runs with those from our Primary set in different $\beta$ ranges. For this purpose we consider all multiples of 0.001 in $\beta$  up to 5.910, where we have high-resolution data for both sets of runs, and multiples of 0.01 in $\beta$ within the interval [5.92, 6.75], where all systems in both sets were run for $10^{10}$ years if they did not become unstable prior to that time. Results are shown in Table~\ref{tab:primary}. The first two rows of results show the averages of the absolute and relative difference in $\log{t_c}$ between the Primary runs and our SL09 runs; subsequent rows detail the fractions of runs within various logarithmic distances of their counterpart at the same value of $\beta$ but different initial longitudes. We find that runs in the Primary Set, which were performed with initial longitudes approaching conjunction, are on average slightly longer-lived than the SL09 longitudes Set, wherein the planets begin at widely-separated initial longitudes.  The bottom row gives the results of a Kolmorogov-Smirnov test of the hypothesis that the ensemble of system lifetimes within the specified range of  $\beta$ for the two Sets were drawn from the same distribution; p-values $< 0.05$ are considered evidence that they were \emph{not} drawn from the same distribution. Note that the comparison is made without considering correlations between $t_c$ and $\beta$ within the specified range of orbital separation. 
 
\section{Chaos}\label{sec:chaos}

We examined the effects of  very small changes in initial longitudes by performing additional integrations with initial conditions identical to our SL09 Set of calculations (\S\ref{sec:SL09}) apart from moving one of the planets along its orbit by 100 meters ($\approx 3.8 \times 10^{-8}$ degrees). As discussed in \S\ref{sec:integrator}, the manner in which  the {\tt Mercury} package reads in longitudes is not accurate enough to make a change this small, so we used the modification generously provided to us by John Chambers that is specified in Footnote 4 for both re-running some of the systems with the nominal SL09 longitudes and performing the slightly different ``Chaos'' integrations. We used resolution of 0.01 in $\beta$ and considered the regions $3.47\le \beta \le 6.32$ and $6.42\le \beta \le 7.12$.  These ranges in $\beta$ were selected because results presented in the previous section suggested that system lifetimes would be short enough that we could run all simulations to orbit crossing, and this proved to be the case. To distinguish the two sets of runs that were performed using this modified code from those presented in Section \ref{sec:SL09}, we refer to them as the SL09$^\dagger$ Set and the Chaos Set. 

\begin{figure}[H]
\centering
\includegraphics[scale=0.75]{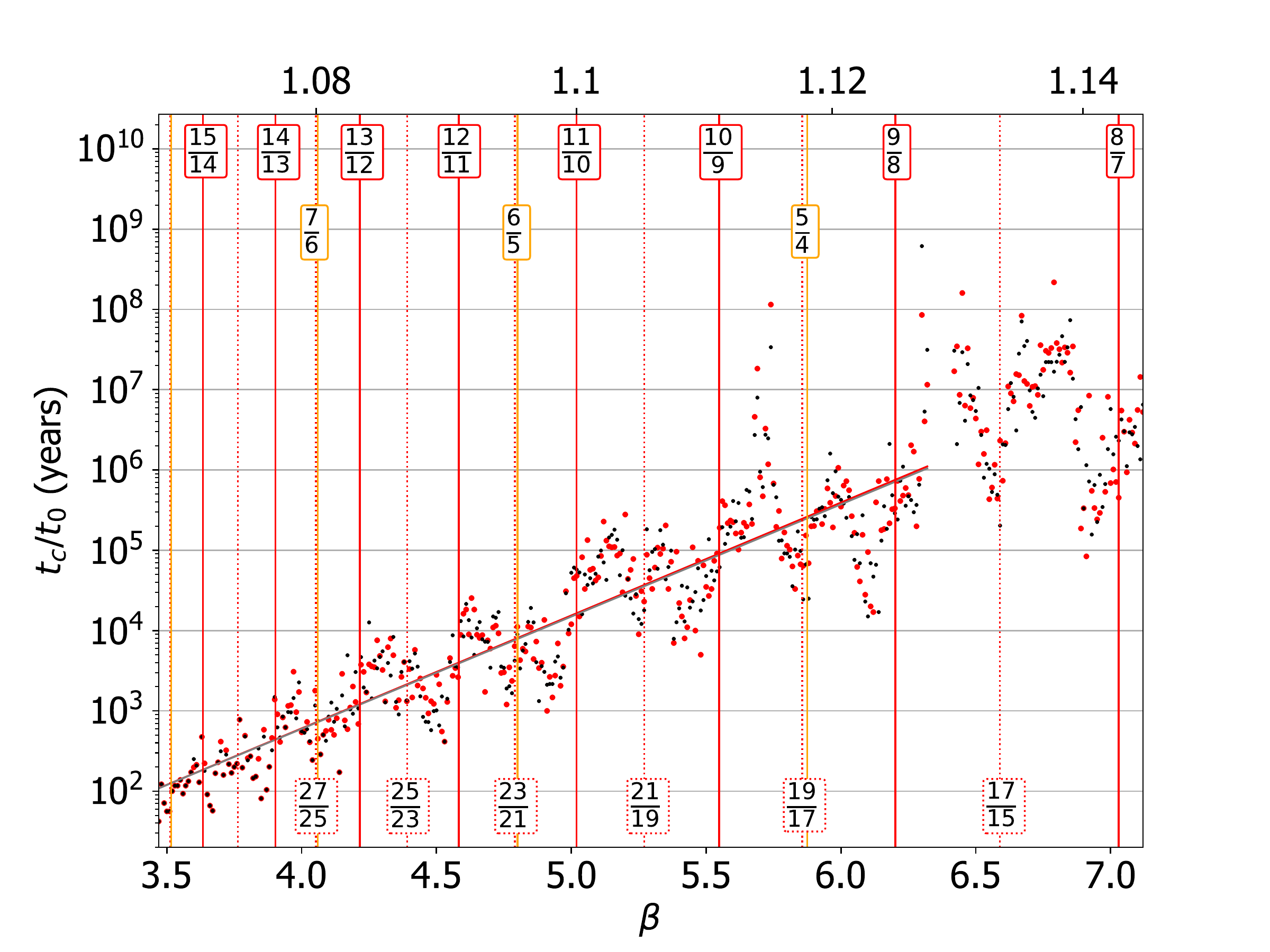}
\caption{Stability time, $t_c$, of systems of three 1~M$_\oplus$ planets orbiting a 1~M$_\odot$ star, displayed as a function
of the initial dynamical separation $\mathrm{\beta}$. Black points are from SL09$^\dagger$ initial conditions. Red points represent systems that were started with the innermost planet displaced by 100 $\mathrm{m}$ ahead in initial longitude. Both sets shown here used the modified code to more accurately input initial longitudes of the planets.  The solid red line corresponds to the fit to the Chaos runs and the solid gray line shows the fit to the SL09$^\dagger$ runs, both over the range $3.47  \le \beta \le 6.32$.} 
\label{fig:chaos}
\end{figure}

Results of our chaos test comparing crossing times of systems with one planet differing in initial longitudes by 100 meters are displayed in Figure \ref{fig:chaos}. Statistical comparisons with lifetimes of systems in SL09$^\dagger$ Set with those of the Chaos systems and the SL09 systems at the same separations are presented in Tables \ref{tab:chaos} and \ref{tab:correction}, respectively.  Differences caused by the change in the code are of comparable magnitude to those caused by moving one planet by 100 meters along its orbit.
Most of the systems that live for less than $1000$ orbits have virtually the same crossing time in all three Sets of integrations. Chaos has more substantial effects (in $\log t_c$) on the lifetimes of longer-lived systems, but except in the $6.00 \le \beta \le 6.32$ range, which includes few runs and thus poor statistics, these tables show a greater fraction of systems with survival times differing by less than a factor of two than for the substantially different longitude combinations compared in Tables \ref{tab:primary}, \ref{tab:alt} and \ref{tab:aligned}.   As expected, the Kolmorogov-Smirnov tests that we performed all were fully consistent with the compared data sets having been drawn from the same populations.

Note that here and elsewhere in this study, we measured the time at which a pair of orbits first crossed.  Had we continued our simulations until an actual collision occurred (which would have required the use of a different integration algorithm), then larger differences would be seen for many of the systems where the crossing times are nearly identical, although these differences do not   increase in proportion to the lifetime of the runs \citep{Bartram:2021}.

\begin{table}[H]
\caption{ Comparison of System Lifetimes: SL09 vs.~Chaos Longitudes}
\scriptsize
\begin{center}
\begin{tabular}{| l || c | c | c | c || c || c |}
\hline
Interval in $\beta$ & \multicolumn{1}{c|}{[3.47, 3.99]} & \multicolumn{1}{c|}{[4.00, 4.99]}& \multicolumn{1}{c|}{[5.00, 5.99]} & \multicolumn{1}{c||}{[6.00, 6.32]} & \multicolumn{1}{c||}{[3.47, 6.32]}  & \multicolumn{1}{c|}{[6.42, 7.12]} \bigstrut \\ \hline
number of runs in the range                                         & 53                    & 100                  & 100                   & 33                       & 286 & 71 \\ \hline
$<\log{t_c}$(SL09$^\dagger$)$-\log{t_c}$(Chaos)$>$                  &  --~0.016              &  --~0.002              & --~0.007            &  --~0.060   & --~0.013 & --~0.044\\ \hline
$<|\log{t_c}$(SL09$^\dagger$)$-\log{t_c}$(Chaos)$|>$                &   0.033             &  0.189              &  0.289               &  0.405         & 0.220  & 0.372 \\ \hline
\# ${t_c}$(Chaos) $<0.5{t_c}$(SL09$^\dagger$)                           & 0 (0.0\%)         & 10 (10.0\%)       & 18 (18.0\%)     &  7 (21.2\%)     & 35 (12.2\%)  & 13 (18.3\%) \\ \hline
\# $0.5{t_c}$(SL09$^\dagger$)$<{t_c}$(Chaos) $<2{t_c}$(SL09$^\dagger$) &  52 (98.1\%)   &  78 (78.0\%)     & 60 (60.0\%)      & 17 (51.5\%)     & 207  (72.4\%) & 40 (56.3\%)  \\ \hline
\# $2{t_c}$(SL09$^\dagger$)$<{t_c}$(Chaos)                               &  1 (1.89\%)        & 12 (12.0\%)     & 22 (22.0\%)      & 9 (27.3\%)     & 44  (15.4\%) & 18 (25.4\%) \\ \hline

\# within 10\% of the SL09$^\dagger$ runs                                   & 39 (73.6\%)     &  21(21.0\%)      & 4 (4.0\%)         & 1 (3.03\%)      & 65 (22.7\%) & 6 (8.45\%)\\ \hline
\# within 1\% of the SL09$^\dagger$ runs                                     & 34 (64.2\%)*     & 6 (6.00\%)*        & 1 (1.00\%)*        & 0 (0.00\%)      & 41 (14.3\%)* & 2 (2.82\%) \\ \hline
Kolmogorov-Smirnov test (p-value) & {0.999} & {0.992}  &  {0.443} & {0.403} & {0.959} & {0.863} \\ \hline
\end{tabular}
 \end{center}
   \tablecomments{ \label{tab:chaos}Comparison of the survival times of integrations for the chaos test (innermost planet 100 m ahead) with those using SL09 longitudes at the same value of $\beta$ using the modified {\tt Mercury} code. The * symbols are explained in the caption to Table \ref{tab:primary}.}
 \end{table}

 \begin{table}[H]
\caption{ Comparison of System Lifetimes: SL09 vs.~SL09$^\dagger$ Longitudes}
\scriptsize
\begin{center}
\begin{tabular}{| l || c | c | c | c || c || c | }
\hline
{Interval in $\beta$} & \multicolumn{1}{c|}{{[3.47, 3.99]}} & \multicolumn{1}{c|}{[4.00, 4.99]} & \multicolumn{1}{c|}{[5.00, 5.99]} & \multicolumn{1}{c||}{[6.0, 6.32]} & \multicolumn{1}{c||}{[3.47, 6.32]} & \multicolumn{1}{c|}{[6.42, 7.12]} \bigstrut \\ \hline
number of runs in the range                                            & 53                    & 100               & 100                & 33                    & 286  & 71\\ \hline
$<\log{t_c}$(SL09)$-\log{t_c}$(SL09$^\dagger$)$>$                  &  0.006              & --~0.011           &  0.069            &  --~0.015   & 0.019 & --~0.018\\ \hline
$<|\log{t_c}$(SL09)$-\log{t_c}$(SL09$^\dagger$)$|>$                &  0.017              &  0.170           &  0.318            & 0.362    & 0.215 & 0.311 \\ \hline

\# ${t_c}$(SL09$^\dagger$) $<0.5{t_c}$(SL09)                           &  0 (0.0\%)        & 7 (7.00\%)    & 25 (25.0\%)    & 10 (30.3\%)    &  42 (14.7\%) & 15 (21.1\%) \\ \hline
\# $0.5{t_c}$(SL09)$<{t_c}$(SL09$^\dagger$) $<2{t_c}$(SL09) & 53  (100.0\%) & 84 (84.0\%)  & 59 (59.0\%)     & 15 (45.4\%)   &   211 (73.8\%) & 40 (56.3\%)\\ \hline
\# $2{t_c}$(SL09)$<{t_c}$(SL09$^\dagger$)                               &  0 (0.0\%)       & 9 (9.00\%)    & 16 (16.0\%)     & 8 (24.2\%)    &  33 (11.5\%) & 16 (22.5\%)\\ \hline

\# within 10\% of the SL09 runs                                   & 47  (88.68\%)     & 23 (23.0\%)  & 4 (4.00\%)      & 3 (9.09\%)      & 77 (26.9\%) & 8 (11.3\%) \\ \hline
\# within 1\% of the SL09 runs                                     & 43 (81.13\%)*     & 9 (9.00\%)*        &  0 (0.00\%)   & 0 (0.00\%)      & 52 (18.2\%)*  & 1 (1.41\%) \\ \hline
Kolmogorov-Smirnov test (p-value)& {0.999} & {0.140}  & {0.556}  & {0.811} & {0.750} & {0.953} \\ \hline
\end{tabular}
 \end{center}
   \tablecomments{ \label{tab:correction}Comparison of the survival times of standard integrations for SL09 longitudes vs.~SL09$^\dagger$ integrations that began at the nominally the same longitudes but using the modified {\tt Mercury} code. The * symbols are explained in the caption to Table \ref{tab:primary}.}
 \end{table}

\section{Hexagonal Longitudes}\label{sec:altlong}

\begin{table}[htbp]
\caption{ Comparison of System Lifetimes: Primary vs.~Hexagonal Longitudes}
\scriptsize
\begin{center}
\begin{tabular}{| l || c | c | c || c || c | c | }
\hline
Interval in $\beta$ & \multicolumn{1}{c|}{[3.465, 3.999]} & \multicolumn{1}{c|}{[4.0, 4.999]} & \multicolumn{1}{c||}{[5.0, 5.91]} & \multicolumn{1}{c||}{[3.465, 5.91]} & \multicolumn{1}{c|}{[6.10, 6.72]} &  \multicolumn{1}{c|}{[8.01, 8.19]} \bigstrut \\ \hline
number of runs in the range &   536 & 1000 & 911 & 2447 & 63 & 19   \\ \hline
$<\log{t_c}$(Primary)$-\log{t_c}$(Hexagonal)$>$          & 0.002 & 0.096 & 0.117 & 0.083 & 0.107 & 0.022   \\ \hline
$<|\log{t_c}$(Primary)$-\log{t_c}$(Hexagonal)$|>$         & 0.307 & 0.325 & 0.400 & 0.352 & 0.349 & 0.298 \\ \hline

\# ${t_c}$(Hexagonal) $<0.5{t_c}$(Primary)&   111 (20.75\%) & 290 (29.00\%) & 251 (27.55\%) & 652 (26.66\%) & 17 (26.98\%) & 4 (21.05\%)  \\ \hline
\# $0.5{t_c}$(Primary)$<{t_c}$(Hexagonal) $<2{t_c}$(Primary) &   306 (57.20\%) & 551 (55.10\%) & 466 (51.15\%) & 1323 (54.09\%) & 36 (57.14) & 11 (57.89\%) \\ \hline
\# $2{t_c}$(Primary)$<{t_c}$(Hexagonal) &   118 (22.06\%) & 159 (15.90\%) & 194 (21.30\%) & 471 (19.26\%) & 10 (15.87\%) & 4 (21.05\%)  \\ \hline

\# within 10\% of the Primary longitude runs & 39 (7.29\%) & 73 (7.30\%) & 78 (8.56\%) & 190 (7.76\%)& 3 (4.76\%) &  3 (15.79\%)\\ \hline
\# within 1\% of the Primary longitude runs  &   4 (0.75\%) &  8 (0.80\%) & 6 (0.66\%)* & 18 (0.74\%)* & 0 (0.00\%) &  0 (0.00\%) \\ \hline
 Kolmogorov-Smirnov test (p-value) &   {0.230}  & {0.001}  & { 0.064} & { 0.063} & { 0.925} & {0.956} \\ \hline
\end{tabular}
 \end{center}
   \tablecomments{ \label{tab:alt}Comparison of the survival times of integrations using Hexagonal longitudes with those at the same value of $\beta$ from our Primary integrations. The resolution is  0.001 over the interval [3.465, 5.910] and 0.01 within the intervals [6.10, 6.72] and [8.01, 8.19]. The * symbols are explained in the caption to Table \ref{tab:primary}.}
 \end{table}
 
\begin{figure}[H]
\centering
\includegraphics[scale=0.75]{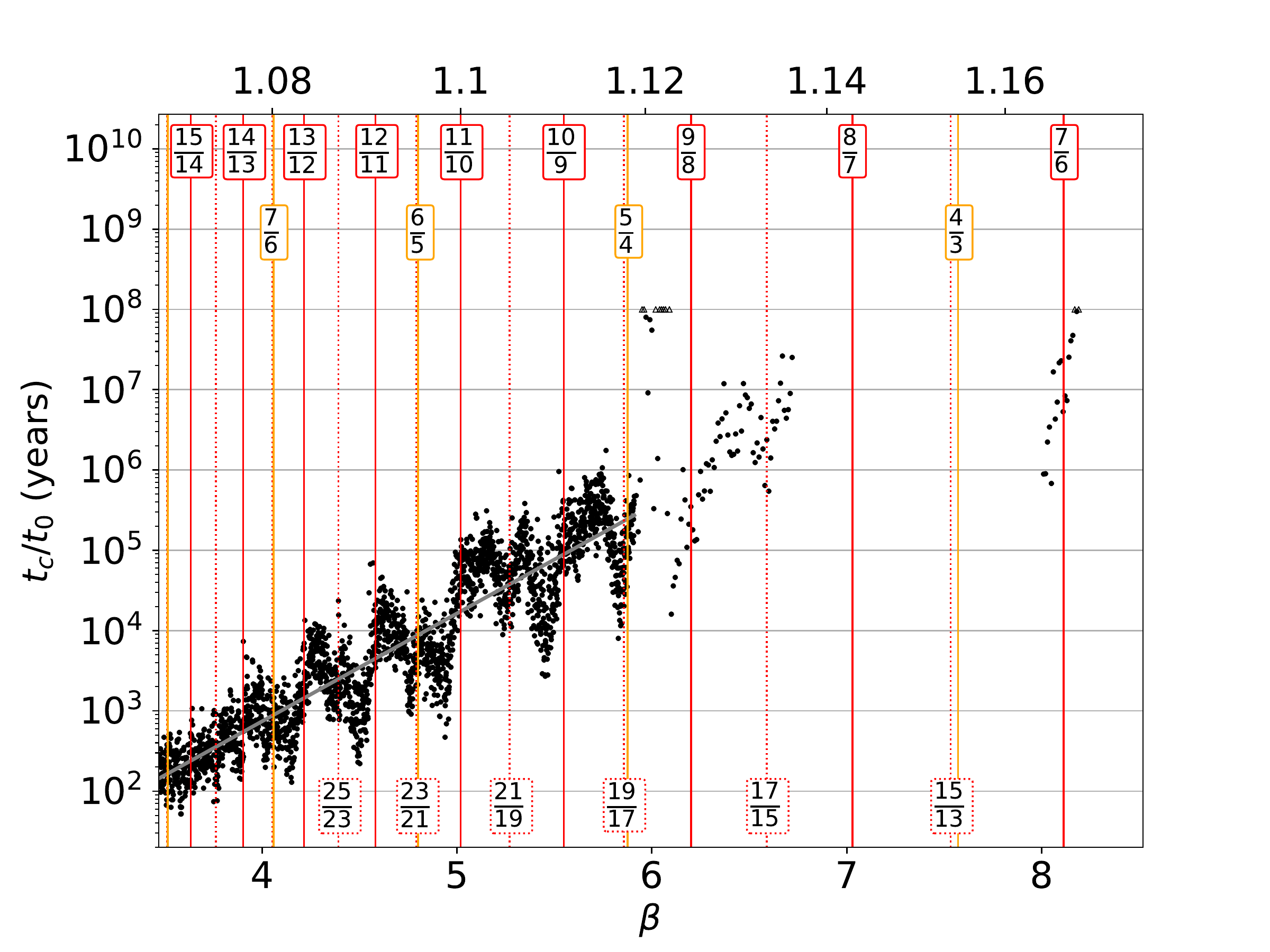}
\caption{{ Survival times of integrations using Hexagonal longitudes plotted against orbital separation, $\beta$. } The black points denote systems with Hexagonal longitudes. The solid gray line corresponds to the fit to lifetimes of systems with Hexagonal longitudes over the range 3.465 -- 5.91 at 0.001  resolution.  }
\label{fig:alt}
\end{figure}

 In order to provide additional data on the sensitivity of the systems' lifetimes to initial longitudes, we simulated a set of systems in which the middle planet started 180$^\circ$ behind the inner planet and the outer planet started 60$^\circ$ ahead of the inner planet; we refer to these as Hexagonal longitudes systems.  These simulations were performed for values of $\beta$ that were multiples of  0.001 over the interval [3.465, 5.91] and multiples of 0.01 within the intervals [5.92, 6.72] and [8.01, 8.2]. All  Hexagonal longitudes integrations were stopped if no orbits had crossed after $10^8$ years; the most tightly-spaced system to survive for $10^8$ years had $\beta = 5.95$. Lifetimes of the systems with these initial longitudes  are shown in Fig.~\ref{fig:alt}.  

Results of statistical studies comparing lifetimes of the systems in the Primary and Hexagonal Sets over various ranges in $\beta$ are shown in Table~\ref{tab:alt}. Systems in our Primary set of runs, which start with the planets approaching conjunction, tend on average to be slightly longer-lived than those in the set with Hexagonal initial longitudes,  with statistically significant differences between the two sets of lifetimes seen over the region $4 \le \beta < 5$ and differences of lesser significance seen for more widely-spaced orbits.

\section{Initially Aligned Longitudes}\label{sec:aligned} 

We perform integrations  for systems with the three planets initially aligned (i.e., at conjunction) in the range [3.465,  5.800] at resolution 0.001 in $\mathrm{\beta}$  and covered  the range [5.81, 7.15] at resolution 0.01. We find 60 systems with $\beta \le 5.8$ that survived at least $\mathrm{10^8}$ years, all having $\beta$ within the range [5.155, 5.354], including 5 within [5.155, 5.187], 10 within [5.188, 5.202], 18 in [5.215, 5.234] and 27 in the interval [5.317, 5.354].  We only integrated the five long-lived systems within the most closely-spaced of these ranges for  $> \mathrm{10^8}$;  four  of these survived longer than $\mathrm{10^9}$ orbits, and two systems  reached $\mathrm{10^{10}}$ years. The most tightly-packed system that survived more than $\mathrm{10^{9}}$ years (it lasted for $\mathrm{6 \times10^{9}}$ years)  has a $\mathrm{\beta}$ value of 5.155, the next one is 5.160 with a crossing time of 1.25 billion years. The two systems that survived for the entire $\mathrm{10^{10}}$ years integrated have  dynamical separations of 5.170 and 5.187.  

Unlike the other sets for fixed initial longitudes that we studied, where the first upward spikes were located in alternating zones between first and second-order resonances, the first  spikes in lifetime for the set of runs with initially aligned longitudes were located in adjacent zones of stability between first and second-order resonances. 

A statistical comparison of the lifetimes of systems more tightly spaced than $\beta = 5.188$ (all of which were integrated until two orbits had crossed or ${10^{10}}$ years had elapsed) with those in our Primary Set of runs is presented in Table \ref{tab:aligned}.  The large magnitudes and negative signs of the numbers representing the mean difference between the logarithms of lifetimes given in the first results row of Table \ref{tab:aligned} (compared to the analogous data in Tables \ref{tab:primary} --  \ref{tab:alt}) imply that initially aligned systems are typically longer-lived than the systems at the other longitude combinations analyzed herein. {\ The Kolmorogov-Smirnov test shows that the computed ensembles of lifetimes are not consistent from being drawn from the same distribution for any of the regions that were compared.}

\begin{figure}[H]
\centering
\includegraphics[scale=0.75]{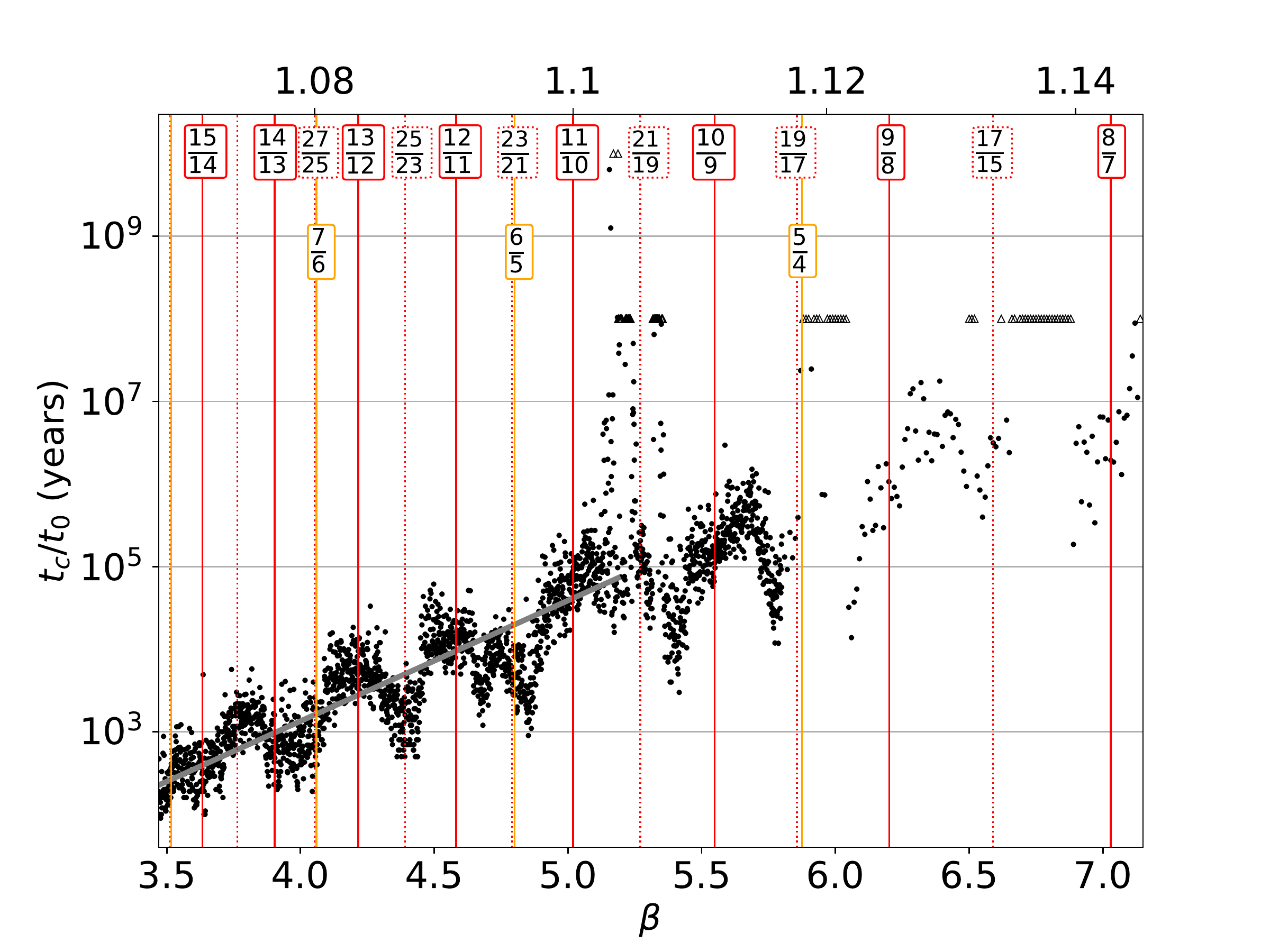}
\caption{Lifetime, $t_c$, of three-planet systems with initially Aligned longitudes as a function of the separation in units of Hill sphere radii $\beta$ (Eq.~\ref{eq:beta}). The solid gray line corresponds to the fit over the whole range over which simulations were allowed to proceed for $10^{10}$ virtual years unless they went unstable,  [3.465, 5.187].}
\label{fig:aligned}
\end{figure}

\begin{table}[htbp]
\caption{ Comparison of System Lifetimes: Primary vs.~Aligned Longitudes}
\begin{center}
\begin{tabular}{| l  r || c | c | c ||  c | }
\hline
Interval in $\beta$ && \multicolumn{1}{c|}{[3.465, 3.999]} & \multicolumn{1}{c|}{[4.000, 4.999]} & \multicolumn{1}{c||}{[5.000, 5.187]} & \multicolumn{1}{c|}{[3.465, 5.187]} \bigstrut \\ \hline
number of runs in the range &  & 535 & 1000 & 188 & 1723   \\ \hline
$<\log{t_c}$(Primary)$-\log{t_c}$(Aligned)$>$        & & --~0.24 & --~0.21 & --~0.36 &  --~0.23   \\ \hline
$<|\log{t_c}$(Primary)$-\log{t_c}$(Aligned)$|>$       && 0.43 & 0.49 & 0.53 & 0.47 \\ \hline

\# ${t_c}$(Aligned) $<0.5{t_c}$(Primary)&&  74 (13.93\%)  & 202 (20.2\%) &  20 (10.64\%) & 296 (17.18\%)  \\ \hline
\# $0.5{t_c}$(Primary)$<{t_c}$(Aligned) $<2{t_c}$(Primary) && 221 (41.31\%) & 384 (38.4\%) & 93 (49.47\%) & 698 (40.51\%)  \\ \hline
\# $2{t_c}$(Primary)$<{t_c}$(Aligned) && 240 (44.86\%) & 414 (41.4\%) & 75 (39.89\%) & 729 (42.31\%)   \\ \hline

\# within 10\% of the Primary longitude runs&& 36 (6.73\%) & 47 (4.7\%) & 15 (7.98\%) & 98 (5.69\%) \\ \hline
\# within 1\% of the Primary longitude runs&& 7 (1.31\%)* & 6 (0.60\%)* & 1 (0.53\%)* & 14 (0.81\%)* \\ \hline
 Kolmogorov-Smirnov test (p-value) &&  9.97 $\times$ 10$^{-19}$  &5.63 $\times$ 10$^{-11}$  & 2.05 $\times$ 10$^{-4}$ & 2.08 $\times$ 10$^{-9}$   \\ \hline
\end{tabular}
 \end{center}
   \tablecomments{ \label{tab:aligned}Comparison of the survival times of integrations using the initially aligned planets with those at the same value of $\beta$ from our Primary integrations.  The * symbols are explained in the caption to Table \ref{tab:primary}.}
 \end{table}

\section{Summary and Discussion}\label{sec:discussion}  

We  integrated the  trajectories of bodies in more than  18,000 planar 4-body systems, each consisting of three 1 M$_\oplus$ planets initially on nested circular orbits about a  1 M$_\odot$ star, until two orbits crossed or a maximum time of $10^8$ or 10$^{10}$ years had elapsed.  We find the same general trend of system lifetime increasing exponentially with orbital separation apart from dips in lifetimes produced by low-order mean-motion resonances found in previous studies, and present parameters of these fits to this trend and measures of characteristic deviations from these fits for various subsets of our integrations in Table \ref{tab:fit}.

\begin{figure}[H]
\centering
\includegraphics[scale=0.75]{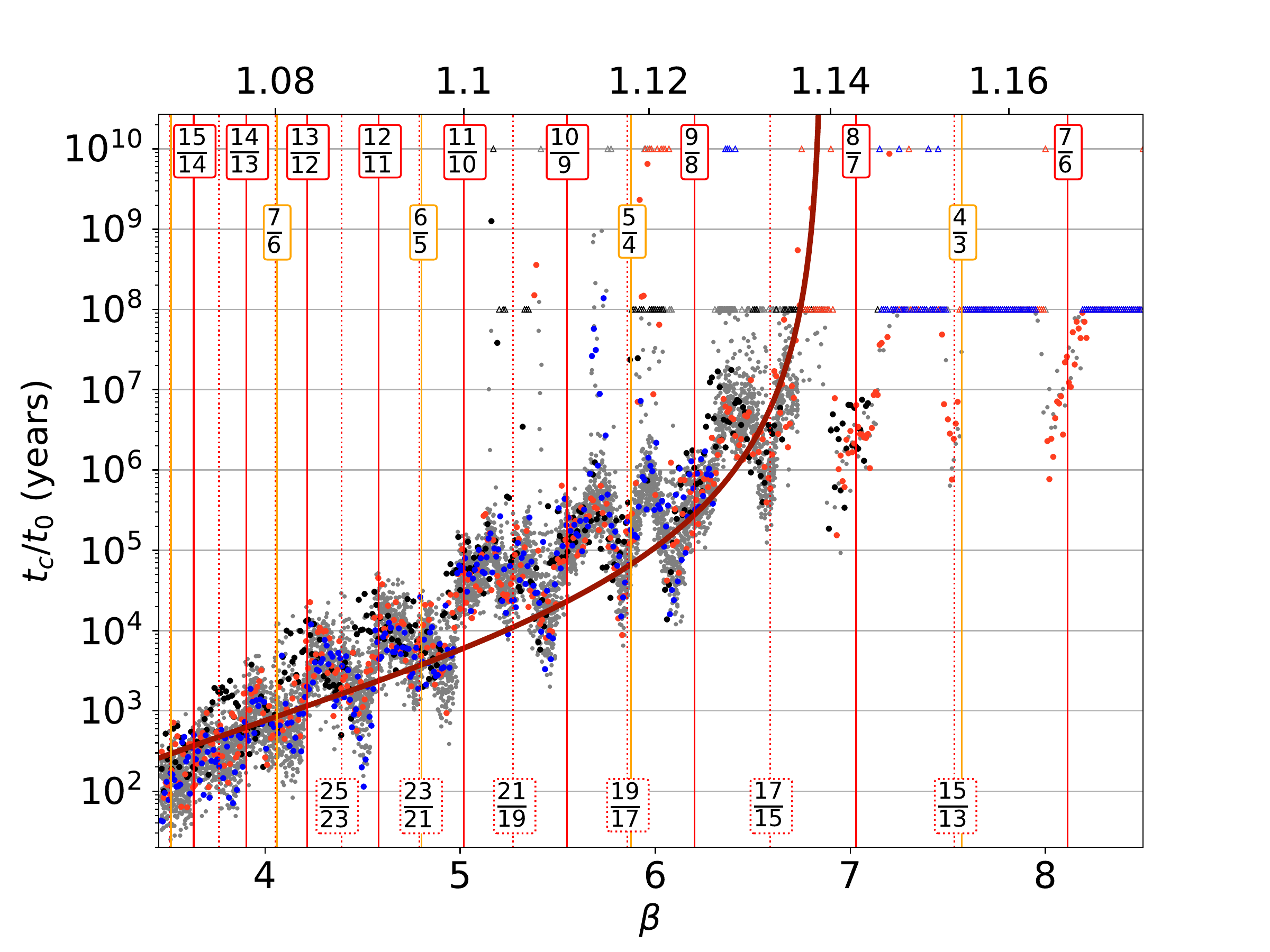}
\caption{Lifetime, $t_c$, of three-planet systems as a function of the separation in units of Hill sphere radii, $\beta$ (Eq.~\ref{eq:beta}), from integrations using various sets of initial longitudes, are reproduced above. Systems with randomly-selected longitudes (\S\ref{sec:OVT}) are represented by small gray dots, those with initial longitudes approaching conjunction (\S\ref{sec:results}) are represented using red dots, those using the same initial longitudes as SL09 (\S\ref{sec:SL09}) in blue, and systems with initially Aligned longitudes (\S\ref{sec:aligned}) in black. Full resolution in orbital spacing is shown for the systems with random longitudes, but only $\beta$ values that are multiples of 0.01 are represented for systems in the other sets. Gray points are plotted first, followed by black,  red and on top blue. The maroon curve shows the analytic results of \cite{Petit:2020}; note the sharp upward slope in the vicinity of $\beta = 6.8$.}
\label{fig:compared}
\end{figure}

The orbital crossing times of the majority of the systems that we integrated are shown in Figure \ref{fig:compared}, together with a curve representing the theoretical prediction of \citet{Petit:2020} for lifetimes of systems with the parameters that we used. The theoretical estimates of stability times are quite good for most of the systems.  However, we found regions of long-term stability for some surprisingly closely-spaced three-planet systems, and destabilization by low-order mean motion resonances disrupts some systems of more widely-spaced planets. 

The most closely-spaced system to survive for the full $10^{10}$ virtual years integrated has initial separation between neighboring orbits of $\beta = 5.17$ mutual Hill radii\footnote{ Although we quote planetary separations in terms of Hill radii for convenience, our results are specific to 1 M$_\oplus$ planets orbiting a  1 M$_\odot$ star (or other combinations with the same planets/star mass ratio). As planetary mass, $M_p$, is varied, single-encounter dynamics scale approximately with Hill sphere size, $(M_p/M_\star)^{1/3}$, whereas two-planet resonance overlap scales as $(M_p/M_\star)^{2/7}$, three-planet resonance overlap scales as $(M_p/M_\star)^{1/4}$, and locations of individual resonances are independent of mass ratio.}, which is less than 50\% wider than the critical separation required for Hill stability of analogous two-planet systems and 25\% closer than the separation beyond which systems of this type are expected to be very long-lived based on the widths and spacings of three-body resonances \citep{Petit:2020}.  The regions of stability near 5.17 and 5.4 mutual Hill radii (\S\ref{sec:aligned}) are narrow and ``porous'' (wide swings in lifetimes with small changes in $\beta$ and/or for differing initial planetary longitudes), implying small volumes in the phase space of initial conditions. But the stable regions beginning at $\beta \approx 5.7$ are far broader, albeit  restricted in longitude and not continuous for $\beta < 7.15$.  We present a more focussed and intensive analysis of very closely-spaced and long-lived three-planet systems in Gavino \& Lissauer (in preparation).

  Two distributions of instability times within spikes are evident in our data (Table \ref{tab:spikes}).  The number of survivors in the spikes for Random longitudes as well as for  the most tightly-packed spike regions for Primary, SL09 and Aligned longitudes show drops by a factor of $\sim 2$ per order of magnitude in time.  The spikes at greater orbital separation in the Primary and Aligned Sets and the one spike seen in the Hexagonal Set have a much larger fraction of long-term survivors. (The statistics are insufficient to assess the SL09 spikes for larger orbital separations.) Perhaps there is something different about the processes driving destabilization within different spikes or different longitudes within a spike; alternatively (or additionally), lifetimes of some of the systems within the wider separation spikes may be orders of magnitude greater than $10^{10}$ years, and the decay rates may become roughly logarithmic at much later times than considered herein.  Note that the first two spikes in the Aligned runs are parts of the same spike, divided into two for inclusion in Table \ref{tab:spikes} because only the inner portion of the region was simulated beyond $10^8$ yr.  Also, for the spike surrounding $\beta = 6$, survivors for Random longitudes give a drop off quickly, whereas Primary, Hexagonal and Aligned longitudes show much more gradual loss of systems with time. Thus, the different rates of dropoff are more likely to represent end members of a continuum rather than distinct distributions.

\cite{Petit:2020}'s estimate for the lifetimes of 3-planet systems with the Earth/Sun mass ratio turns up sharply near period ratio 1.138 (Fig.~\ref{fig:compared}), beyond which lifetimes of many systems are expected to be much longer because three-body resonances no longer overlap \citep{Quillen:2011,Petit:2020}. This period ratio corresponds to $\beta \approx 6.8$ in our Hill sphere radius normalized units,  placing it within the last longitude-limited upward spike in lifetimes (moving a little farther out is the dip in lifetimes associated with the 8:7 MMR of neighboring planets, followed by the first upward spike present at all longitude combinations examined).    

 The broad extents in initial orbital separation and limited ranges of starting planetary longitudes for the spikes in system lifetime observed for $5.5 < \beta < 7$  provide strong evidence that the long survival times of these systems are \emph{not} the result of planets librating within protective resonances. These spikes extend for large enough ranges in orbital separation that the differences in the relative longitudes of the planets at various $\beta$ values within a spike circulate many times within the number of orbits given by the general exponential trend in $t_c(\beta)$ near the spike. This suggests that the phasing of the early conjunctions of the planets can, in a nontrivial fraction of the cases, have profound effects on the lifetimes of three-planet systems.  

\begin{figure}[H]
\centering
\includegraphics[scale=0.50]{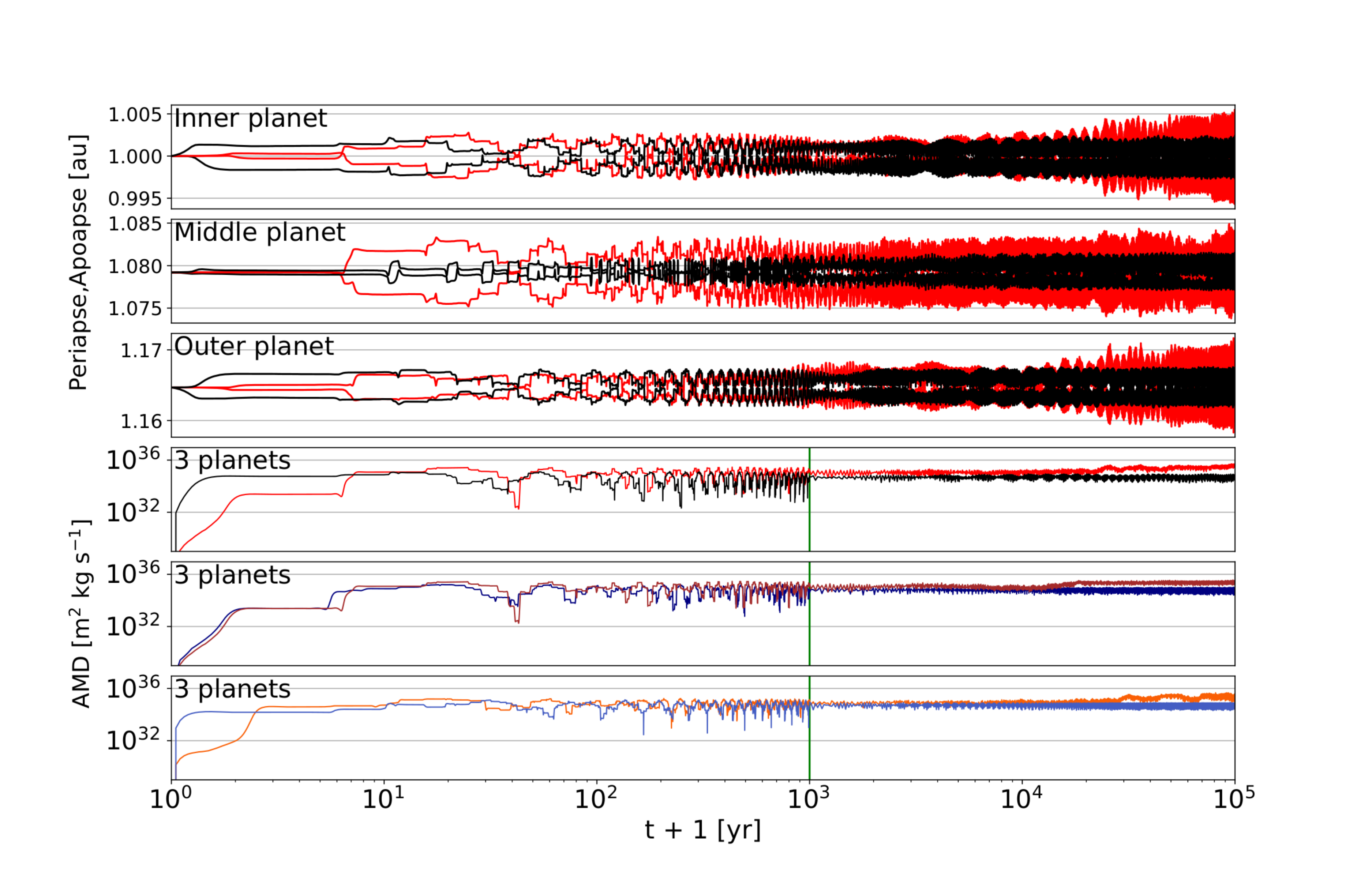}
\caption{ Evolution of six planetary systems with $\beta = 6.02$ over the first $10^5$ years. One year has been added to the time in order to plot the evolution of the system from the start with a logarithmic time scale. Top three panels: Radial distance (in au) of the periapses and apoapses for each of the  three planets in the Primary run (black; plotted on top) and the SL09 run (red; hidden behind the black curve where the two curves overlap). The fourth panel shows the  AMD (Eq.~\ref{eq:AMD}) of these systems in S.I. (MKS) units (note that this panel and the ones below it are log-log plots) with the same plotting convention, except that for $t > 1000$ years, we show AMD averaged over 15 years to reduce overlap between the two curves. The fifth panel displays the AMD of the Hexagonal longitudes system (navy blue) plotted on top of the curve for the Chaos system (brownish red), and the bottom panel compares the AMD of the long-lived Aligned system (royal blue) with that of the short-lived Random longitudes run (orange).  }
\label{fig:qQplot}
\end{figure}

In our formalism, all planets begin on circular orbits; we do not try to minimize quantities analogous to proper eccentricities. Figure \ref{fig:qQplot} compares the early evolution of  six of our planetary systems with $\beta = 6.02$; three of these systems, those with Primary, Hexagonal  and Aligned initial longitudes, are very long lived ($t_c > 10^8$ years), whereas the systems  beginning with the SL09, Chaos   and Random longitudes each go unstable in less than $10^6$ years.  For our Primary longitudes, where the planets begin shortly before a three-planet (near) conjunction, eccentricities $e \approx 0.0014$ are imparted to the inner and outer planets during the initial planetary conjunction occurring in the first orbit. But the first encounter of the middle planet with its neighbors in the system starting at SL09 longitudes, which occurs in the eighth orbit, together with subsequent planetary encounters during the following two decades, boost the planetary eccentricities. Therefore, the angular momentum deficit (AMD) of this system, which for planar orbits is given by the formula 
\begin{equation}
\label{eq:AMD}
\mathrm{AMD} = m\sqrt{G M a}\big(1 - \sqrt{1 - e^2}\big),
\end{equation}

\noindent grows quickly to values far higher than those obtained by the planets started at the Primary longitudes. The initial conjunctions are different for our systems beginning at our Hexagonal  and Aligned longitudes, but the low AMD of the system compared to the values for the Chaos  and Random longitudes systems is also established within the first few synodic periods.

 Table \ref{tab:fit} presents four measures of the scatter of system lifetimes within each range in orbital separation and Set of runs for which we computed log-linear fits on data with resolution of 0.01 in $\beta$ or better. In most cases, both measure of dispersion relative to the fit lines, $\sigma_\mathrm{exp}$ and   $\sigma_\mathrm{exp}^\mathrm{cum}$, grow larger as the upper limit to the range of orbital separations increases. Also, for a given range in orbit separation, both measures of scatter about the local trend, $\sigma_\mathrm{local}$ and   $\sigma_\mathrm{local}^\mathrm{cum}$, are typically larger for the Set of runs with randomly-chosen initial longitudes than for Sets of systems with the same combination of initial longitudes, indicating a significant dependance of lifetimes on the starting longitudes of the planets.

Deterministic chaos accounts for some of the scatter in lifetime seen among the systems with similar initial conditions that we integrated, but its effects are typically small, especially for relatively short-lived systems (Fig.~\ref{fig:chaos}). Note that we only followed systems until two planetary orbits had crossed. Much larger chaotic variations, especially in fractional changes of the lifetimes of short-lived systems, are found if the orbits are followed to physical collisions of planets with parameters analogous to those of Earth \citep{Bartram:2021}.

\section{ Comparison With Previous Numerical Studies}\label{sec:compare}

\citet{Tamayo:2016} studied a grid of nonuniformly-spaced systems of three 5 M$_\oplus$ planets; the spacings of the inner pair, $\beta_i$, and the outer pair, $\beta_o$, were selected randomly and independently within the range [5, 9]; eccentricities and inclinations were drawn randomly over [0, 0.02] and [0, $1^\circ$], respectively. They did not find any system (of $\sim 5000/16\approx 310$ simulated) with both $\beta_i < 6$ and $\beta_o < 6$  that was stable for $10^7$ years, and only 5 of the $\sim 625$ with  $\beta_i + \beta_o < 12$ endured for the entire $10^7$ initial inner planet orbits that they allowed their integrations to run (Daniel Tamayo, private communication, 2020).   They use more massive planets, a difference that is expected to increase stability for the majority of $\beta$ values since resonance overlap (which varies as the 2/7 power of the planets/star mass ratio, \citet{Wisdom:1980}) and 3-body resonance density (proportional to the 1/4 power of the planets/star mass ratio, \citet{Quillen:2011}) both scale with smaller powers of mass than the Hill sphere size (1/3 power of the planets/star mass ratio). Their inclusion of nonzero $e$ acts in the opposite sense \citep{Gratia:2021}; the small mutual inclinations of the planets that they integrate probably have little effect on average.  \citet{Petit:2020} did not find any systems of three $\sim~3$~M$_\oplus$ planets more tightly-spaced than both of their analytic stability estimates that survived for the full $10^9$ years that they simulated; they did, however, find a few closely-spaced systems of roughly Mercury-mass planets that survived for the entire 1 Gyr.

\cite{Obertas:2017} (OVT17) integrated systems that included five 1 M$_\oplus$ planets  with randomly and independently drawn values of $\beta$ (corresponding to an average resolution of 0.0005 for $\beta < 10$) and initial longitudes. The most-closely spaced system to survive for the entire $10^{10}$ years that they simulated had $\beta = 8.6397$ (Alysa Obertas, private communication, 2020), corresponding to 2.4941 times the critical separation for Hill-Jacobi stability of two-planet systems.  

Our results for three-planet systems  are, in many respects, qualitatively similar  to those found for analogous 
five-planet systems analyzed by OVT17. Lifetimes of closely-packed, uniformly-spaced, three-planet systems and five-planet systems each typically increase exponentially with distance between the orbits. In most of the regions of phase space that we examined, system lifetimes depend primarily on orbital separation, with starting angular separations between the planets being far less important, in agreement with the findings of OVT17 for five-planet systems. 

 But there are significant differences between the dynamics of three-planet systems and of five-planet systems.  Three-planet systems are generally longer-lived, with the magnitude of the difference increasing at wider separations, in agreement with the findings of \citet{Chambers:1997},  \citet{Smith:2009} (SL09), and \citet{Funk:2010}.  A more qualitative difference that was not previously recognized is that closely-spaced three-planet systems have narrow  regions of parameter space that allow for stability for orders of magnitude longer than most systems with similar orbital separations. These regions are located far from strong mean motion resonances. This likely results from three-planet systems having much smaller numbers of combinations of two planets (3 vs.~10) and of three planets (1 vs.~10), resulting in fewer strong two-planet and three-planet mean motion resonances.  

 For separations $\beta < 7.1$ (slightly more than twice the separation required for Hill stability of two planets), there are no regions in $\beta$ where spikes are present in both of the initial planetary longitude combinations that we studied most extensively (Sections \ref{sec:results} and \ref{sec:SL09}).   This is qualitatively different from the spikes present in OVT17's data for 5-planet systems (all with $\beta > 8.5$), where system lifetimes have little dependence on initial planetary longitudes, which were selected randomly and independently for each of the systems that they integrated. Thus, the dynamics of systems of three planets are profoundly different from those of systems with five (or, presumably, more) planets.

Resonances can destabilize planetary systems, but they can also enhance stability. However, in the systems studied herein, as well as those integrated by SL09 and OVT17,  the only clear signature of resonances in the lifetime plots is destabilization.  Regions of stable resonance trappings for three or more planets in known systems such as TRAPPIST-1 are small \citep{Tamayo:2017}. In resonance chains, superperiods\footnote{The superperiod of two planets near but not in a first-order MMR is equal to the periodicity of the near-resonant argument. Its value is given by $|2\pi[Nn_2-(N-1)n_1]^{-1}|$.} are equal, and for small $e$ cases such as Kepler-80 and TRAPPIST-1, superperiods are short enough that period ratios for pairs near the same first-order resonance increase nontrivially as one moves outwards (a few parts per thousand from one planet pair to the next), whereas in our formalism, the superperiods for the inner and outer pairs are nearly equal (compare the black and green points in Fig.~\ref{fig:average}).  Thus, not only is the stable region of parameter space small, for small planetary eccentricities it is also somewhat displaced from the region in which all of these studies have searched.

The zones of stability for $\beta < 6.75$, and especially for $\beta < 5.65$, occupy small fractions of parameter space (Table \ref{tab:spikes}).  However, these islands of stability are large enough to be found by a survey of parameter space, as opposed to the minuscule stable fractions of parameter space  that are occupied by some actual planetary systems with planets in resonance-lock librations \citep{Tamayo:2017}, which have only been reproduced numerically by dissipating systems into resonances.

  Values of our measure of deviations of system lifetimes relative to the median lifetime of systems with similar orbital separations, $\sigma_\mathrm{local}
^\mathrm{cum}$,  listed in Table \ref{tab:fit} for cases where the resolution in orbital separation is 0.001 or better are similar to or slightly larger than dispersion measures found for shadow trajectories of planar 5-planet systems by \cite{Hussain:2020} and those for nearly planar 4-planet systems given by \cite{Rice:2018}. This suggests that much of the local dispersion is the result of chaos, with lesser contributions of systematic variations in lifetimes caused by very small changes (differences of $\le 0.005$ mutual Hill radii) in initial orbital separation.

\citet{Quarles:2018} performed stability studies of systems of 2 -- 6  planets (each 1~M$_\oplus$) orbiting in the habitable zones of $\alpha$ Centauri A and $\alpha$ Centauri B, with perturbations of the companion star being taken into account. For a given magnitude of characteristic system lifetime, their figures show substantially more scatter in lifetime for systems of two planets (and thus four massive bodies, like our systems of three planets orbiting a single star) than for systems of three or more planets.  However, their calculations were not at sufficiently high resolution in orbital separation to determine whether or not spikes of extremely large increases in lifetime are present for  two-planet systems around either of the stars in this nearby stellar binary.

\section{ Epilogue: Relativity, Tides, and Planets on Short-Period Orbits}\label{sec:epi}

The integrations reported upon in this work were run until either a pair of planetary orbits crossed or the elapsed time reached a pre-determined value of either $10^8$ or $10^{10}$ times the initial orbital period of the inner planet ($10^8$ or $10^{10}$ years for the inner orbit located 1 AU from a 1~M$_\odot$ star).  Ten billion year integrations are sufficient for 1~M$_\oplus$ planets in the habitable zone of a sunlike star, as the Universe is only slightly older than this, and moreover the main sequence lifetime of a  1~M$_\odot$ star is $\sim 10^{10}$ years.  However, the dynamical age (number of orbital periods of the inner planet) of most \ik multiple planet systems is $10^{11}-10^{12}$ orbits, so one may ask why have we (and almost all previous studies of this type) limited our simulations to dynamical ages of at most $10^{10}$ inner planet orbits, and are the results of such time-limited simulations relevant for the stability of \ikt's multi-planet systems? 
 
As noted in Section \ref{sec:methods}, our integrations used purely Newtonian dynamics of point mass particles, neglecting general relativistic deviations (GR) as well as tidal forces, both of which are more important for planets orbiting very close to their stars. Neglecting GR and tides makes the problem scale-invariant, allowing our results to be applied to planetary systems orbiting closer to their stars and to lower (higher)-mass planets orbiting lower (higher)-mass stars provided orbits and masses were varied by the same factor for all bodies in the system\footnote{Analogous scalings applicable to planetary accretion simulations that include physical collisions are given in Appendix C of \citet{Quintana:2006}.}. In contrast, including GR and/or tides would specialize the study to specific planetary (and stellar) masses and orbital sizes, voiding the scalings discussed at the beginning of \S \ref{sec:methods} and expanding the dimensionality of the parameter space of the (already computationally-demanding) problem. Incorporating tidal damping would require even more specialization than accounting for GR because of the great diversity of tidal response characteristics that terrestrial planets may possess.

Although the first-order post-Newtonian deviations from classical dynamics caused by general relativity are important in stabilizing our Solar System, this is because GR moves the locations of resonances rather than making fundamental changes to the dynamics \citep{Batygin:2008, Laskar:2008}. Therefore, GR may well not be important to statistical samples appropriate for understanding general scalings for planetary systems except in the regime where it substantially alters orbital precession rates or some other key variable.

Tides are more fundamental to the problem than are the principal post-Newtonian terms in GR, as they induce dissipative processes that fundamentally alter the dynamics of the system\footnote{Gravitational radiation also removes mechanical energy from the system, but the amount lost is exceeding small for all known multi-planet systems.}, e.g., by damping eccentricities that have been excited by perturbations among the planets. Tidal damping is likely to be strong enough to prevent the applicability of dissipation-free simulations to planets with orbital periods of a few weeks or less for timescales of billions of years, but tidal damping  drops off rapidly with increasing orbital distance. Thus, given the age of the Universe, non-dissipative integrations (the type presented herein) lasting for more than $10^{11}$ orbits would be of minimal applicability to exoplanet systems.  Integrations for $10^{10}$ orbits require wall clock times of several days using modern cpu cores, so integrating for longer than this would severely reduce the number of systems that it would be feasible to integrate. Although our longest integrations are only one-tenth of the desired duration of $10^{11}$ orbits, instabilities tend to occur logarithmically with time, i.e., the range in $\beta$ for which orbits cross increases logarithmically with time. The fundamental timescale for instabilities is the synodic period of the planets, which is of order ten orbits for the closely-spaced systems considered herein.  Thus, in a logarithmic sense, results for integrations for $10^{10}$ orbits would be expected to reveal roughly 90\% of the physics that occurs within $10^{11}$ orbits. Eventually, as computing costs continue to drop, it will be feasible to test this hypothesis with a statistically-significant number of $10^{11}$ orbit integrations.

\section{Acknowledgments}
John Chambers kindly provided us with the code to improve the accuracy of input angles for the {\tt Mercury} integration package (see Footnote 4 for details). We thank Pete Bartram for pointing out an error in the listed initial longitudes of planets in some of the simulations that appeared in an early draft of this paper.   Rus Belikov, Tony Dobrovolskis and Dan Tamayo provided constructive comments on a draft of our manuscript. JJL was supported through NASA's PSD ISFM program.

\end{document}